\documentclass[10pt,conference]{IEEEtran}
\usepackage[T1]{fontenc}
\usepackage[utf8]{inputenc}
\usepackage[english]{babel}
\usepackage[style=ieee, maxnames=9, minnames=9, date=year, doi=false, isbn=false, url=false, backend=biber, sortcites]{biblatex}
\usepackage{nicefrac}
\usepackage{mathtools,amsthm,amssymb,amsfonts}
\usepackage{csquotes}
\usepackage{graphicx}
\usepackage{subcaption}
\usepackage{booktabs}
\usepackage[detect-all]{siunitx}
\usepackage[usenames,dvipsnames]{xcolor}
\usepackage{microtype}
\usepackage[noend]{algorithm,algpseudocode}
\floatname{algorithm}{Algorithm}
\usepackage{hyperref}
\usepackage[compat=0.6]{yquant}

\usetikzlibrary{calc, positioning, fit}

\makeatletter
\let\oldtheequation\theequation
\renewcommand\tagform@[1]{\maketag@@@{\ignorespaces#1\unskip\@@italiccorr}}
\renewcommand\theequation{(\oldtheequation)}
\makeatother

\newtheorem{example}{Example}

\newcommand*{\ket}[1]{\ensuremath{|#1\rangle}}
\AtBeginDocument{}

\newcommand*{\qop}[1]{\ensuremath{\textit{#1}}}
\newcommand{\cnot}{\qop{CNOT}}
\newcommand{\cex}{\qop{CEX}}
\newcommand{\csum}{\qop{CSUM}}

\hypersetup{
	pdftitle={Mixed-Dimensional Quantum Circuit Simulation with Decision Diagrams},
	pdfsubject={QCE 2023},
	pdfauthor={Kevin Mato, Stefan Hillmich, Robert Wille} 
}

\begin{document}

\title{Mixed-Dimensional Quantum Circuit Simulation with Decision Diagrams}
\author{
\IEEEauthorblockN{%
	Kevin Mato\IEEEauthorrefmark{1}\hspace{1cm}%
	Stefan Hillmich\IEEEauthorrefmark{2}\IEEEauthorrefmark{3}\hspace{1cm}%
	Robert Wille\IEEEauthorrefmark{1}\IEEEauthorrefmark{2}%
	}
\IEEEauthorblockA{\IEEEauthorrefmark{1}Chair for Design Automation, Technical University of Munich, Munich, Germany}\IEEEauthorblockA{\IEEEauthorrefmark{2}Software Competence Center Hagenberg (SCCH) GmbH, Hagenberg, Austria}
\IEEEauthorblockA{\IEEEauthorrefmark{3}Institute for Integrated Circuits, Johannes Kepler University Linz, Linz, Austria}

\IEEEauthorblockA{\href{mailto:kevin.mato@tum.de}{kevin.mato@tum.de}, \href{mailto:stefan.hillmich@scch.at}{stefan.hillmich@scch.at}, \href{mailto:robert.wille@tum.de}{robert.wille@tum.de}}%
\IEEEauthorblockA{\url{https://www.cda.cit.tum.de/research/quantum/}\vspace*{-1em}}
}

\maketitle
\begin{abstract}
    Quantum computers promise to solve several categories of problems faster than classical computers ever could.
    Current research mostly focuses on qu\emph{b}its, i.e.,~systems where the unit of information can assume only two levels.
    However, the underlying physics of most (if not all) of the technological platforms supports more than two levels, commonly referred to as qu\emph{d}its.
    Performing computations with qudits increases the overall complexity while, at the same time, reducing the number of operations and providing a lower error rate. 
    Furthermore, qudits with different number of levels can be mixed in one system to ease the experimental control and keep representations as compact as possible. 
    Exploiting these capabilities requires dedicated software support to tackle the increased complexity in an automated and efficient fashion.
    In this paper, we present a qudit simulator that handles mixed-dimensional systems based on \emph{Decision Diagrams} (DDs).
    More precisely, we discuss the type of decision diagram introduced as underlying data structure as well as the resulting implementation.
    Experimental evaluations demonstrate that the proposed solution is capable of efficiently simulating mixed-dimensional quantum circuits, with specific use cases including more than 100 qudits in one circuit. 
    The source code of the simulator is available via \href{https://github.com/cda-tum/MiSiM/}{github.com/cda-tdum/MiSiM} under the MIT~license.
\end{abstract}


\section{Introduction}
\label{sec:introduction}
Quantum computing utilizes a different computing paradigm that promises to solve several categories of problems compared to classical computers.
Examples include Shor's algorithm~\cite{shor} to factorize integers, Grover's search~\cite{grover1996fast} for unstructured data, and evaluating possible materials as catalysts in quantum chemistry~\cite{Cao_2019}. 
Around the globe, many research groups in academia and industry (such as Google, IBM, and Microsoft) try realize this potential and push on what technology can achieve today.

However, thus far, the considered applications where mostly limited to systems composed of two-dimensional qubits.
This neglects a large potential available through systems of higher dimensions, inherently possessed by almost any underlying technology of physical realizations of quantum computers. 
The usage of higher dimensions increases the overall complexity of building circuits but it also enables compression of previously costly non-local gates between qubits into local gates on a single qu\emph{d}it---increasing the fidelity of the resulting state~\cite{ringbauer2021universal}.

The abstract idea for higher-dimensional systems and corresponding theory has been around for quite some time~\cite{wang2020qudits}.
Fundamentally, qudits enable denser storage of information and provide a much larger set of possible operations compared to qubits.
Given these advantages, basic control has been demonstrated in physical platforms such as trapped ions~\cite{Zhang2013Contextuality,ringbauer2021universal}, 
to photonic systems~\cite{Lanyon2008,Ringbauer2017Coherence, Hu2018a, Malik2016}, 
\mbox{superconducting circuits}~\mbox{\cite{Kononenko2020,Morvan2020}}, 
Rydberg atoms~\cite{Ahn2000}, 
nuclear spins~\cite{Godfrin2017}, 
cold atoms~\cite{Anderson2015}, 
nuclear magnetic resonance systems~\cite{Gedik2015} and molecular spin~\cite{molecularspin}. 

Recent developments in quantum algorithms have shown that multi-level logic is a more natural architecture for implementing complex applications~\cite{deller2022quantum, RydbergSim}. Simulations of models representing fermion-boson interactions on mixed-dimensional quantum computers could enable real-time simulations of quantum electrodynamics and other field theories with continuous or larger symmetry groups~\cite{quditbasedQED, phitheory, fermBoson}.
Gate decompositions on \mbox{mixed-dimensional} systems offer reduced complexity due to the temporary expansion of the Hilbert space~\cite{shortcuts}. This leads to smaller circuits and a higher chance of success due to less noise accumulation. Optimizing the usage of qudits of different dimensions further improves circuit compactness and error rate, as seen in recent research.

However, given these recent breakthroughs from physicists, there is the risk of an emerging \emph{design gap}, where powerful multi-dimensional quantum computers are available but we do not have means to utilize their power.
To avoid this situation, dedicated methods and software support are required to keep up in the design automation domain before the point is reached, where manually designing systems, circuits, and controls is not tractable anymore.

We contribute to that by presenting a classical simulator for mixed-dimensional quantum circuits, i.e.,~circuits where each qudit may have a different dimensionality. To this end, we propose an extension to edge-weighted Decision Diagrams~\cite{Miller2006,DBLP:conf/iccad/ZulehnerHW19,DBLP:journals/tcad/ZulehnerW19} which serve as the main data-structure to cope with the (exponential) complexity of simulation. 
In the past, decision diagrams have been proven to be a suitable data structure to compactly represent exponentially-sized data in many cases~\cite{DBLP:conf/date/AbdollahiP06,DBLP:journals/ieicet/WangLTK08,DBLP:books/daglib/0027785,Miller2006,DBLP:journals/tcad/NiemannWMTD16,DBLP:conf/iccad/ZulehnerHW19}.
The type of decision diagram proposed in this work dynamically captures the dimensionality of each qudit in the \mbox{mixed-dimensional} system to minimize the required memory.
Further, it elegantly visualizes the individual dimensionalities of the qudits.
The experimental evaluation on a set of benchmarks confirms the efficacy of the simulator (publicly available at \href{https://github.com/cda-tum/MiSiM}{github.com/cda-tum/MiSiM}) as tool for design automation in mixed-dimensional systems.

The remainder of this paper is structured as follows.
\autoref{sec:background} provides the necessary background on quantum states and operations for higher-dimensional systems, i.e.,~qudits.
\autoref{sec:motivation} motivates the problem of quantum circuit simulation and details the contributions of this paper.
\autoref{sec:state-of-the-art} gives an overview of the state of the art in quantum circuit simulation.
\autoref{sec:decision-diagrams} describes the proposed type of decision diagram for mixed-dimensional system and \autoref{sec:implementation} details the corresponding implementation.
\autoref{sec:results} summarizes the experimental evaluation.
Finally, \autoref{sec:conclusions} concludes the paper.

\section{Background}
\label{sec:background}
In this section, we briefly review the basics of quantum information processing with a focus on mixed-dimensional quantum logic and how these concepts scale to the abstraction of quantum circuits.

\subsection{Quantum Information Processing}
\label{sec:QIP}

In classical computing, \emph{bit} (binary digits) are the primary unit of information, which can only exist in either the 0 or~1 state. In quantum computing, \emph{qubits} (quantum bits) are the corresponding unit of information. The key difference from classical computing is that qubits can exist in any linear combination of $\ket{0}$ and $\ket{1}$ (using Dirac's bra-ket notation~\cite{DBLP:books/daglib/0046438}). However, constructing qubits involves restricting the natural multi-level structure of the underlying physical carriers of quantum information. 

Therefore, these systems natively support \mbox{multi-level logic} with the fundamental unit of information termed a \emph{qudit} (quantum digit).
A qudit is the quantum equivalent of a \mbox{$d$-ary} digit with $d\geq 2$, whose state can be described as a vector in the $d$-dimensional Hilbert space $\mathcal{H}_d$. 
The state of a qudit can thus be written as a linear combination $\ket{\psi} = \alpha_0 \cdot \ket{0} + \alpha_1 \cdot \ket{1} + \ldots + \alpha_{d-1} \cdot \ket{d-1}$, or simplified as vector $ \ket{\psi} = \begin{bsmallmatrix} \alpha_0 & \alpha_1 & \ldots & \alpha_{d-1}\end{bsmallmatrix}^\mathrm{T}$, where $\alpha_i \in \mathbb{C}$ are the amplitudes relative to the orthonormal basis of the Hilbert space---given by the vectors $\ket{0}, \ket{1},\ket{2},..., \ket{d-1}$.

The squared magnitude of an amplitude $|\alpha_i|^2 $ defines the probability with which the corresponding basis state $i$ will be observed when measuring the qudit. 
Since the probabilities have to add up to $1$, the amplitudes have to satisfy $\sum_{i=0}^{d-1} |\alpha_i|^2 = 1$.

Two key properties that distinguish quantum computing from classical computing are superposition and entanglement.
A qudit is said to be in a \emph{superposition} of states in a given basis when at least two amplitudes are non-zero relative to this basis. 
\emph{Entanglement}, on the other hand, describes a form of superposition born from interactions in multi-qudit systems. Entanglement is a powerful form of quantum correlation, where the quantum information is encoded in the state of the whole system and cannot be extracted from the individual qudits anymore.

\begin{example}\label{ex:state}
    Consider a system of one qudit with only three energy levels (also referred to as \emph{qutrit}).
    The quantum state $\ket{\psi} = \sqrt{\nicefrac{1}{3}}\cdot\ket{0} + \sqrt{\nicefrac{1}{3}}\cdot\ket{1} + \sqrt{\nicefrac{1}{3}}\cdot\ket{2}$ is a valid state with equal probability of measuring each basis. 
    Equivalently, the quantum state may be represented as vector $\sqrt{\nicefrac{1}{3}}\cdot \begin{bsmallmatrix} 1 & 1 & 1\end{bsmallmatrix}^\mathrm{T}$.
    
    In a similar fashion, quantum systems of \emph{mixed dimensions} can be constructed.
    Extending the previous qutrit state by a qubit enables representation of the following entangled state \mbox{$\ket{\psi'} = \sqrt{\nicefrac{1}{3}}\cdot\ket{0}_3\ket{0}_2 + \sqrt{\nicefrac{1}{3}}\cdot\ket{1}_3\ket{1}_2 + \sqrt{\nicefrac{1}{3}}\cdot\ket{2}_3\ket{0}_2$}---equivalently represented by the vector $\sqrt{\nicefrac{1}{3}}\cdot \begin{bsmallmatrix} 1 & 0 & 0 & 1 & 1 & 0\end{bsmallmatrix}^\mathrm{T}$.
\end{example}

\subsection{Quantum Operations}
The state of a single $d$-level qudit system can be manipulated by operations which are represented in terms of \mbox{$d \times d$-dimensional} unitary matrices $U$, i.e.,~matrices that satisfy $U^\dagger U = U U^\dagger = I$.
This property makes quantum operations logically reversible.
Quantum operations can be divided in two categories: local and non-local--entangling operations. 
Common examples of local operations are the Pauli operations
\begin{equation}
\label{eq:pauli}
        X = 
        \begin{bmatrix}
            0 & 0 & 1 \\
            1 & 0 & 0 \\
            0 & 1 & 0 \\
        \end{bmatrix} \quad
        Z = 
        \begin{bmatrix}
            1 & 0 & 0 \\
            0 & \omega & 0\\
            0 & 0 & \omega^2\\
        \end{bmatrix},
\end{equation}
with $\omega = e^{\nicefrac{2\pi i}{d}}$ and $d$ being the dimension of the single qudit. 
The local operations shown in \autoref{eq:pauli} are the generalization of the qubit operations to a multi-level systems, in this particular case a \emph{qutrit}.

Qudit systems can be entangled and, for this reason, they support a set of non-local operations, although these last operations cannot be simply derived by generalizing entangling qubit operations. Realizing entangling operations in qubit systems is comparatively simple and rather unclear in qudit systems. 
More precisely, for qubit systems it is sufficient to compile the $\cnot$ gate to the native operations of the quantum hardware, because all qubit entangling operations can be implemented by the controlled-\qop{NOT} ($\cnot$) gate and appropriate local operations on the subsystems~\cite{DBLP:books/daglib/0046438}.
In contrast, for qudit systems this does not hold anymore and entanglement can be generated in many in-equivalent ways. 
Consequently, while any single entangling gate is sufficient for universal quantum computation~\cite{brennen2005efficient}, not all entangling gates are equally useful for any given application.

\begin{example}\label{ex:entangling-gates}
Consider, the controlled-exchange gate $\cex$~\cite{ringbauer2021universal} applied on two qudits and defined by
\begin{align}
    \cex_{c,t_1,t_2} : 
    \begin{cases}
        \text{Swap}(t_1,t_2) & \text{ if control is $c$} \\
        \text{Identity} &\text{ otherwise}
    \end{cases}.
    \label{eq:cex}
\end{align}
This qudit-embedded version of the $\cnot$ gate generates qubit-level entanglement in a high-dimensional Hilbert space. However, there are gates that directly generate qudit entanglement, such as the controlled-SUM gate defined by
\begin{align}
    \csum: \ket{c,j} \mapsto \ket{i,i\oplus j} ,
    \label{eq:csum}
\end{align}
where $\oplus$ denotes addition modulo the dimension $d$.
\end{example}

These are just two examples of a more general theme in qudit systems, where entangling gates differ in their \emph{entangling power}. The reasons why non-local qudit gates produce more entanglement than qubit ones are presented in depth in Ref.~\cite{nativequdit}.

\subsection{Quantum Circuit}
After discussing how operations are represented for qudits and mixed-dimensional systems, it is relevant to discuss how to apply a quantum operation, called gates in the quantum circuit representation, and how algorithms for quantum systems can be represented as well. The matrices representing quantum operations describe the evolution of a quantum system. For this reason, the application of a $U$ operator can be determined by multiplying the corresponding input state from the left with the matrix~$U$, and the output state is the final state of the evolution imposed by $U$.

\begin{example}
    \label{ex:mult}
     Consider a three-level qudit (i.e.,~a qutrit) initially in the state $\ket{0}$. Applying the Hadamard operation~$H_3$ to it yields the output state shown before in \autoref{ex:state}, i.e.,
    \begin{align*}H_3\cdot \ket{0}=
        \frac{1}{\sqrt 3}
        \begin{bmatrix}
            1 & 1 & 1 \\
            1 & e^{\frac{2\pi}{3}} & e^{\frac{-2\pi}{3}} \\
            1 & e^{\frac{-2\pi}{3}} & e^{\frac{2\pi}{3}}
        \end{bmatrix} \cdot
        \begin{bmatrix}
            1 \\
            0 \\
            0 
        \end{bmatrix} 
        = \frac{1}{\sqrt 3}
        \begin{bmatrix}
            1 \\
            1 \\
            1 
        \end{bmatrix}.
    \end{align*}
\end{example}

Given multiple qudits of different dimensions and several operations applied to them according to an algorithm, we can illustrate the sequence as a \emph{quantum circuit}.
For a quantum circuit consisting of multiple mixed-dimensional qudits, the unitary matrix will be of dimension $\prod_i d_i \times \prod_i d_i$ with $d_i$ denoting the dimension of each qudit. The application of the operation will require an input state of the size of $\prod_i d_i$ entries. Later on, adjustments to the matrices to match the size of the system are going to be discussed in more depth.

\begin{figure}[tpb]
    \centering
    \includegraphics[width=0.6\linewidth]{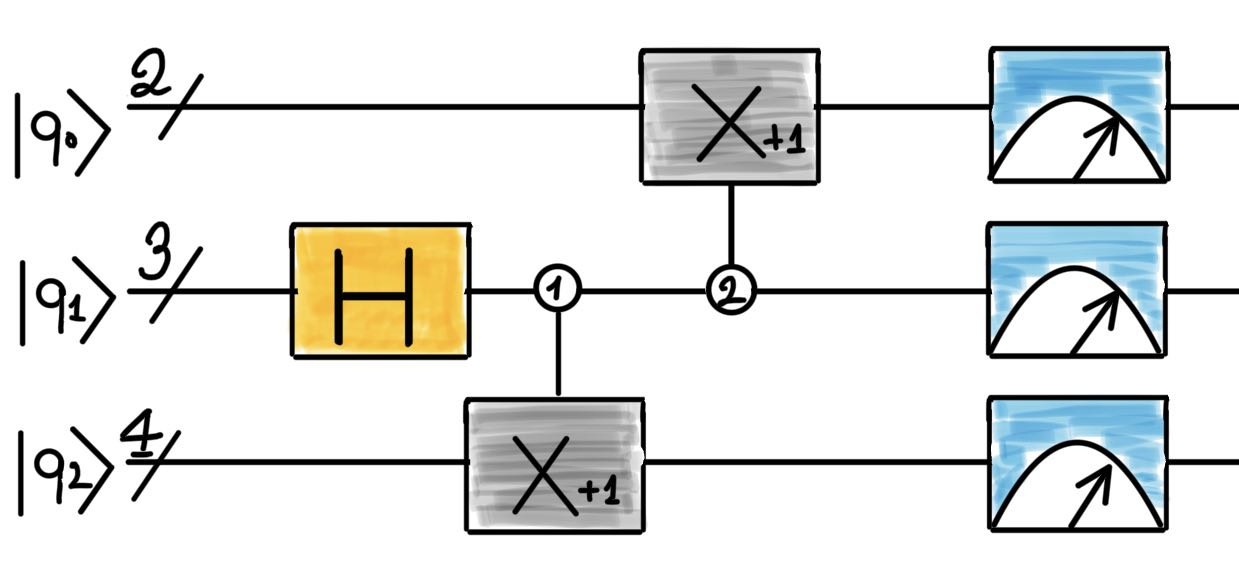}
    \caption{A mixed-dimensional circuit with three qudits}
    \label{fig:mixed_circuit}
\end{figure}


\begin{example}
The quantum circuit in \autoref{fig:mixed_circuit} shows three lines, in order: a qubit, a qutrit (3 levels), and a ququart (4 levels), as well as three different types of operations, a local operation, two controlled operations, and measurements. The controlled operations on qudits apply a unitary to a target line only if the control qudit is in a $\ket{i}$ between $0$ and $d-1$, with $d$ dimensions of the control qudit. The control state is marked inside the circle on the control line. The first operation is a Hadamard applied to the qutrit. Afterwards a controlled-on-1 Pauli X (normally addressed as \cnot) is applied to the ququart and lastly to the qubit. Finally, the three units of information are measured in order to derive the outcomes of the circuit. The state of the system after the application of each operation can be viewed as \emph{intermediate}, as the output state of one is the input state of another.
\end{example}

\section{Motivation}
\label{sec:motivation}
In this work, we investigate how to efficiently simulate quantum circuits where the qudits can have different dimensions, referred to as mixed-dimensional systems.
To this end, the previous section provided the basis for understanding quantum information, with a particular focus on multi-level quantum logic, i.e.,~systems with both qubits and qudits. 
In this section, we review the design automation task of quantum simulation that is going to enable the utilization of mixed-dimensional systems. 
Following that, we discuss the challenges of simulating quantum computations and the benefits of simulating quantum circuits of mixed dimensionality.

\subsection{Simulation}
\label{sec:simulation}

Simulation of quantum circuits is an important task in the domain of design automation.
In essence, the simulation of a quantum circuit is conducted by successively applying operations in sequence to an initial quantum state, i.e.,~multiplying matrices and vectors.
While quantum circuit simulation is simple, the exponential memory requirements in the number of qubits make it hard in practice.
With qudits of higher dimensions, this memory requirement becomes even higher.

More precisely, given a system with $N$ qudits of possibly mixed dimensions, a vector describing this system will have length $D = \prod_{i=0}^{N-1} d_i$, with $d_i$ denoting the dimension of each qudit.
Correspondingly, the matrices representing the operations on this system will have dimension $D\times D$.
In fact, even for operations that affect only a subset of qudits, they have to be \enquote{padded} with the appropriate identity operations to ensure the correct (large) size (using the Kronecker product~\cite{DBLP:books/daglib/0046438}).

\begin{example}\label{ex:exponential}
    Consider the Hadamard gate on a qutrit as shown in \autoref{fig:mixed_circuit}, also denoted $H_3$, affecting the second of the three qudits. 
    Applying this operation leaves the first and third qudit untouched, i.e.,~it applies the identity operation of appropriate dimension.
    This is achieved by combining the Hadamard operation and identity operations via the Kronecker product as in the following illustration:
    \begin{align*}
        I_{2} \otimes H_{3} \otimes I_{4} =
            \scalebox{0.90}{\mbox{\ensuremath{\left[\arraycolsep=1.3pt\def\arraystretch{0.8} \begin{array}{c c c @{\hspace{6pt}} c c c @{\hspace{9pt}} c c c @{\hspace{6pt}} c c c}
                :: & ::  & :: & :: & ::  & :: & :: & ::  & :: & :: & ::  & :: \\
                :: & ::  & :: & :: & ::  & :: & :: & ::  & :: & :: & ::  & :: \\
                :: & ::  & :: & :: & ::  & :: & :: & ::  & :: & :: & ::  & :: \\[3pt]
                :: & ::  & :: & :: & ::  & :: & :: & ::  & :: & :: & ::  & :: \\
                :: & ::  & :: & :: & ::  & :: & :: & ::  & :: & :: & ::  & :: \\
                :: & ::  & :: & :: & ::  & :: & :: & ::  & :: & :: & ::  & :: \\[6pt]
                :: & ::  & :: & :: & ::  & :: & :: & ::  & :: & :: & ::  & :: \\
                :: & ::  & :: & :: & ::  & :: & :: & ::  & :: & :: & ::  & :: \\
                :: & ::  & :: & :: & ::  & :: & :: & ::  & :: & :: & ::  & :: \\[3pt]
                :: & ::  & :: & :: & ::  & :: & :: & ::  & :: & :: & ::  & :: \\
                :: & ::  & :: & :: & ::  & :: & :: & ::  & :: & :: & ::  & :: \\
                :: & ::  & :: & :: & ::  & :: & :: & ::  & :: & :: & ::  & :: 
                \end{array}\right]}}}
    \end{align*}
    Here, each dot represents an element of the combined operation.
    The result is a matrix with $24 \times 24$ entries. 
\end{example}

\autoref{ex:exponential} should convey an intuition about the rapid increase in size for state vector and operation matrices with respect to the number of qudits and their respective dimensionality.
Consequently, the simulation of quantum computations requires an asymptotically exponential amount of memory with respect to the number of qudits.
Single measurements of a quantum state suffer from the same complexity, but repeated measurements can be done in amortized linear time~\cite{DBLP:conf/dac/HillmichMW20}.

The memory complexity can be lowered quite significantly for many cases, especially for common single-qudit operations.
In essence, the operation can be applied to the state vector directly, e.g.,~by swapping rows for the exchange-operation.
Further, quantum states with many zero entries may be represented by sparse vectors~\cite{10.1145/3491248}.
However, the efficacy of this approach diminishes with operations affecting an increased number of qudits and quantum states that are not sparse.



\begin{example}
    Consider the \emph{GHZ state} of two qutrits, i.e.,~qudits of dimension three, as an example of an entangled state, given as $\ket{\psi} = \frac{1}{\sqrt{3}}(\ket{00}$ + $\ket{11}$ + $\ket{22})$.
    The corresponding state~vector is $\psi = \frac{1}{\sqrt{3}}\begin{bsmallmatrix} 1 & 0 & 0 & 0 & 1 & 0 & 0 & 0 & 1\end{bsmallmatrix}$.
    Here, only three out of nine entries are non-zero.
\end{example}

\subsection{The Case for Quantum Circuit Simulation}

As highlighted in the previous section, the simulation of quantum circuits is conceptually simple but not easy in its execution.
This where \emph{design automation} comes into play to enable realizing the impact of high-mixed-dimensional systems.
There is a great need for new automated methods, software frameworks, and theory, where only preliminary results have been achieved~\cite{9926318,DBLP:conf/aspdac/MatoRHW23,DBLP:conf/qce/MatoRHW22, ZXW,mato2023compression}.
In this section, we make the case for utilizing the domain knowledge of design automation to overcome the exponential complexity in many cases.
In fact, the need for better solutions to quantum simulation lies not only in the capability of understanding quantum computations but in enabling the study of new applications and new features created by using mixed-dimensional systems. 

There are three immediate advantages in having an appropriate simulator for mixed-dimensional quantum systems.
\begin{enumerate}
    \item Getting otherwise opaque information about the quantum state.
    \item Enabling design exploration.
    \item Aiding in verification.
    \item Identifying potential for compression.
\end{enumerate}

The inner processes of a quantum computer are fundamentally opaque.
Trying to \enquote{look at} a quantum state will inadvertently make it decay into a basis state -- immediately destroying any superposition or entanglement.
However, we can accurately calculate the exact information of a quantum state by classical means.
Of course, this is only possible for smaller systems due to the exponential overhead in the number of qudits.
Nevertheless, classical simulation is an important tool for algorithm development and debugging.
This becomes even more important when targeting higher- and mixed-dimensional systems.
Their increased complexity also effects the developers of these system, making automated software support a necessity to reduce errors and speed up the design.


The second motivation is that simulation can take on the role of a tool for design exploration.
In fact, in qudit systems, it is overall possible to explore new trade-offs between using higher-levels in the single qudit and using ancilla lines. 
The first one is similar to the effect of using ancilla qubits, but with a simpler circuit complexity and hardware design~\cite{shortcuts}. 
On mixed-dimensional systems this can be achieved with a temporal expansion of the Hilbert space. This leads to smaller circuits that have a higher chance of succeeding due to the smaller noise accumulated. 
However, just increasing the dimension of all qudits in a quantum circuit still leaves room for improvement~\cite{commtrade, timeeff}.
A simulator enables the study of different circuits, composed of more or less ancilla lines, depending on the limitations and potential of the implementing platform. This can lead to a more accurate study of suitable circuits for dedicated platforms and applications.

Classical simulation of quantum circuits also aids in verification schemes.
On the one hand, it helps circuit equivalence checking by simulating two circuits with identical initial states and comparing the output.
This is an easier task than constructing the matrices and comparing these.
Such schemes that incorporate quantum circuit simulation have been proposed~\cite{DBLP:conf/dac/BurgholzerW20,DBLP:journals/tcad/BurgholzerW21}

Finally, it is possible to see through simulations that the computation may happen only on specific levels of a qudit. 
This could allow the compression of a single qudit's computational space from a higher dimension to a lower dimension restricted to only the useful levels~\cite{volyaaspdac}. 
By this, the representations become more compact, reflecting into a competitive error rate and experimental control, typical of the state of art of qudit~systems~\cite{ringbauer2021universal}.
The simulation of \mbox{mixed-dimensional} systems opens new possibilities in the field of circuit compression.
Two examples illustrate the ideas.

\begin{figure}[tpb]
	\centering
	\includegraphics[width=0.8\linewidth]{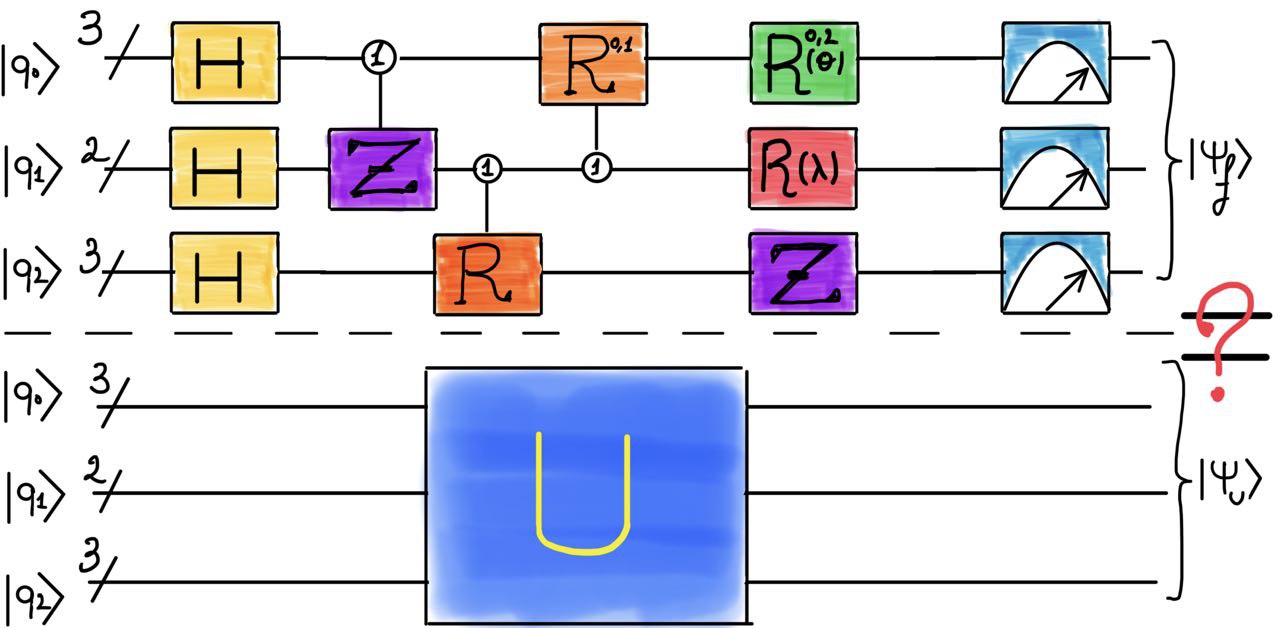}
	\caption{Verification of a quantum circuit given the original algorithm, or the unitary representing the evolution. }
	\label{fig:verify}
\end{figure}

\begin{example}\label{ex:verification}
	The use of a simulator could be of use for verifying that a compiled quantum circuit, a new decomposition, or a particular encoding preserves the intended original functionality defined in the algorithm. In \autoref{fig:verify}, this is illustrated using a simple use case, where a unitary has been compiled to a \mbox{mixed-dimensional} system and the simulation can tell us if, given an initial input state, the final states of the two are identical. If true, the circuit performs correctly the intended algorithm.
\end{example}
\begin{example}\label{ex:necesary}
    Consider the following operation $U$:
    \begin{align}\label{subrotation}
        \setlength\arraycolsep{2pt}
        U = \begin{bmatrix}
        1 & 0 & 0 & 0 & 0  \\
        0 & e^{\frac{2\pi}{3}} & e^{\frac{-2\pi}{3}} & 0 & 0  \\
        0 & e^{\frac{-2\pi}{3}} & e^{\frac{2\pi}{3}}  & 0 & 0 \\
        0 & 0 & 0 & 1& 0 \\
        0 & 0 & 0 & 0 & 1  \\
        \end{bmatrix} .
    \end{align}
    It is possible to note that the local operation is operating only on two levels of the original five-level system (i.e.,~of a ququint). 
    Taken into account that many operations are applied to only a subset of the available levels, we can consider the original system to be of only two dimensions, and consequently remap the logic levels to \ket{0} and \ket{1}. 
    This results in a more compact operation of dimension $2\times 2$, namely
    \begin{align}\label{subrotation2}
        \setlength\arraycolsep{2pt}
        U' = \begin{bmatrix}
            e^{\frac{2\pi}{3}} & e^{\frac{-2\pi}{3}}\\
            e^{\frac{-2\pi}{3}} & e^{\frac{2\pi}{3}}\\
        \end{bmatrix} .
    \end{align}
\end{example}

\subsection{Contribution}

Motivated from the above, we present a \mbox{mixed-dimensional} quantum circuit simulator, based on decision diagrams, with a publicly available implementation. 
The aim is to investigate a new method for an efficient simulation of circuits with qudits of arbitrary and mixed dimensionality.

The proposed approach is motivated by the success of the decomposition schemes used in~\cite{ Miller2006,DBLP:journals/tcad/ZulehnerW19}, that fall under the family of \emph{Decision Diagrams} (DDs;~\cite{DBLP:conf/date/AbdollahiP06,DBLP:journals/ieicet/WangLTK08,DBLP:books/daglib/0027785,Miller2006,DBLP:journals/tcad/NiemannWMTD16,DBLP:conf/iccad/ZulehnerHW19}) used in various fields of design automation, especially in simulation, verification, and synthesis.
Further, the structure of directed acyclic graphs, which resemble tree, can elegantly illustrate the dimensionalities, interactions, and entanglement in mixed-dimensional systems.

This will unlock the potential benefits of this generalized type of quantum circuit by exploiting more compact representations for quantum states and operations on mixed-dimensional systems. 
In the following sections, we provide a review of the current state of art in quantum circuit simulation, introduce decision diagrams and their benefits, and describe in detail our proposed representations and required manipulation algorithms.

\section{State of the Art}
\label{sec:state-of-the-art}

There has been an ongoing and significant interest from researchers and engineers in developing solutions for simulating quantum computations. 
There are several available solutions that individually make use of different data structures, e.g.,~arrays~\cite{qiskit,quest,qibo,yao}, decision diagrams~\cite{DBLP:journals/tcad/ZulehnerW19,DBLP:books/daglib/0027785}, tensor networks~\cite{qsimCirq,tenpy}, and matrix-product states~\cite{tenpy}.
However, the vast majority of these simulators have been developed so far with the focus on qubit systems and their related circuits, while qudit systems have not been provided with the same software support.
Qubit simulators can be divided in two main categories.

The first category of simulators is referred to as \mbox{\emph{array-based}}.
These simulators have limitations due to their reliance on the straightforward representation of quantum states and operations. 
Typically, simple 1-dimensional and 2-dimensional arrays are used. 
To simulate larger quantum systems, they require massive hardware power, such as HPC infrastructures composed of thousands of nodes, and petabytes of distributed memory. 
The utilization of accelerators and dedicated hardware can improve the run-time of the matrix-vector multiplications.

The second category of simulators relies on specialized data structures.
For example, solutions based on decision diagrams~\cite{DBLP:journals/tcad/ZulehnerW19,DBLP:books/daglib/0027785}, have been proposed to exploit redundancies to gain a more compact representation of state vectors and matrices. 
Simulators based on tensor networks, such as those found in popular software packages ~\cite{quimb, itensor}, share a similar goal of simulating quantum systems. However, they achieve this goal by compressing the information using a network of tensors that are contracted together in a specific way. One example of such a simulation method is the \emph{Matrix Product States} (MPS), which efficiently calculates physical properties of \mbox{one-dimensional} quantum systems~\cite{tenpy}. There are also examples of simulators using a combination of \mbox{matrix-vector} multiplication and tensor networks as qsim~\cite{qsimCirq}, which supports a variety of circuit models. 

However, the available implementations offer preliminary support of higher-dimensional systems at best, if at all.
There are early examples of trials to qudit simulations of quantum circuits of homogeneous dimensions, that peaked with a \mbox{proof-of-concept} prototype of QMDDs~\cite{Miller2006}. 
More recent attempts come from the established framework for quantum circuits manipulation and simulation, Google's Cirq~\cite{qsimCirq}. 
It provides the core functionalities for the simulation of quantum circuits but requires the scientists to explicitly integrate the software with the desired gate set and special circuit constructions. 
Hence, unfortunately, Cirq supports mixed-dimensional quantum circuit simulation only in principle.
The recently published qudit simulator QuDiet~\cite{qudiet} provides a qubit-qudit quantum simulator package with quantum circuit templates of well-known qubit quantum algorithms for fast prototyping and simulation. 
QuDiet shares several similarities with Cirq and qsim~\cite{qsimCirq} in terms of software structure and philosophy. 
So far, the existing simulators are focused on providing interfaces for the development of specific use cases rather than being ready-to-use qudit simulators. 
This lack of availability of the current software platforms and subsequent lack of automated methods is a major obstacle in development for higher- and mixed-dimensional systems.

\section{Decision Diagrams}
\label{sec:decision-diagrams}

A key aspect of developing an efficient simulator is to conquer the correspondingly resulting exponential complexity for as many cases as possible.
In the past, decision diagrams have been proven to enable efficient representation of exponentially-sized data in many cases~\cite{DBLP:conf/date/AbdollahiP06,DBLP:journals/ieicet/WangLTK08,DBLP:books/daglib/0027785,Miller2006,DBLP:journals/tcad/NiemannWMTD16,DBLP:conf/iccad/ZulehnerHW19,DBLP:journals/tcad/ZulehnerW19,DBLP:conf/qce/HillmichHRMW21,Wille2023}.
In essence, they achieve their compactness by exploiting redundancy in data they represent.
This section introduces the decision diagrams for representing quantum states and operations of mixed-dimensional systems.

A \emph{Decision Diagram} (DD) is a \emph{Directed Acyclic Graph} (DAG), composed of nodes and directed edges.
The nodes are organized in levels, which each level representing one qudit.
The edges represent the connection between the qudits and have additional information annotated to them.
This construction can represent quantum states and quantum operations as detailed in the following sections.

\subsection{Quantum States}

The general idea of using decision diagrams to represent quantum states is based on repeated decompositions of the corresponding vectors.
To this end, consider an $n$-qudit quantum state $q_{n-1}, q_{n-2}, \ldots, q_0$, where $q_{n-1}$ is arbitrarily designated the \enquote{most sigificant qudit}.
This state is split into equally-sized parts based on the dimensionality of qudit $q_{n-1}$.
Each part is represented by a successor node which also encodes the decision for the value of $q_{n-1}$.
The splitting procedure is then repeated for each part until the individual complex entries in the original vector are reached.
During this process, the complex amplitudes of each basis state are stored in the edge weights and equal sub-vectors (up to a complex factor) are represented by the same nodes.
To guarantee canonicity, the nodes are normalized such that the sum of squared magnitudes of out-edges of a node add up to one.
The resulting decision diagram has a \emph{root node} $q_{n-1}$, at least one node for each further $q_i$, and finally a single \emph{terminal node}, which does not have any successor.
The decisions on the value of the qudits is recorded so it can be reconstructed.
Reconstructing the amplitude for individual basis states is achieved by accordingly traversing the decision diagram and multiplying all edge weights along the path.
An example illustrates the idea.

\begin{figure}[tpb]
    \centering
    \includegraphics[width=0.55\linewidth]{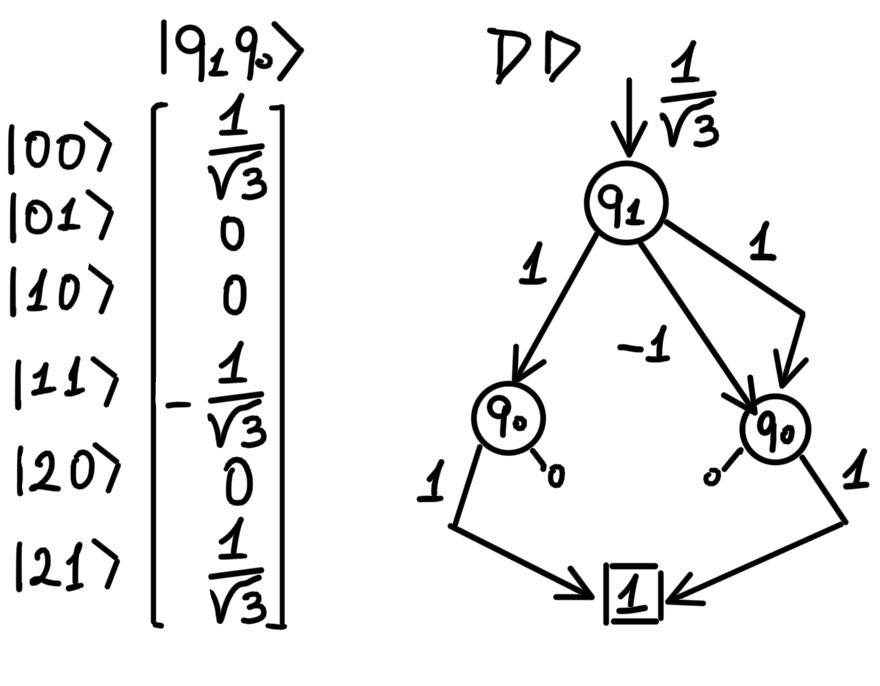}
    \caption{A state vector of a qutrit-qubit entangled state and the corresponding DD representation.}
    \label{fig:statedd}
\end{figure}
    
\begin{example}
    Consider the quantum state $\frac{1}{\sqrt{3}}(\ket{00} - \ket{11} + \ket{21})$ in a qutrit-qubit system.
    \autoref{fig:statedd} gives both, the representation as vector and as decision diagram.
    The vector has 6 entries due to the product of the local dimensionalities: $3$ for the qutrit and $2$ for the qubit. 
    The root node $q_1$ has 3 outgoing edges, one for each possible level in qutrit. 
    The nodes in the second level represent the qubit and have 2 edges each. 
    The second and the third edge of the root node point to the same qubit node, due to the exploitation of redundancy.
    The $q_0$ nodes point to the \emph{terminal node}.
    
    Finally, in order to retrieve an amplitude of a basis state, we start by multiplying the weights along the path composed of the basis state we want to retrieve. 
    In the of case $\ket{00}$, we compose $\nicefrac{1}{\sqrt{3}} \cdot 1 \cdot 1$, corresponding to the weights of the root node, the edge $0$ of $q_1$, and the edge $0$ of $q_0$. 
    In case of the bitstring \ket{11}, we multiply $\nicefrac{1}{\sqrt{3}} \cdot -1 \cdot 1$, corresponding to the weights of the root node, the edge $1$ of $q_1$ and the edge $1$ of~$q_0$.
\end{example}

\subsection{Quantum Operations}

The representation of quantum operations as decision diagram follows a similar scheme to that of quantum states.
Instead of splitting the vector, for quantum operations the corresponding matrix is split into equal parts depending on the dimenionality of the \enquote{most significant qudit}, e.g.,~$3 \times 3 = 9$ parts in case of a qutrit.
This scheme is again applied until the individual complex numbers in the matrix are reached.
Equal sub-parts up to a complex constant are recognized and stored by the same node in the decision diagram to achieve compactness and canonicity.
Again, an example illustrates the idea.

\begin{figure}[tpb]
    \centering
    \includegraphics[width=0.8\linewidth]{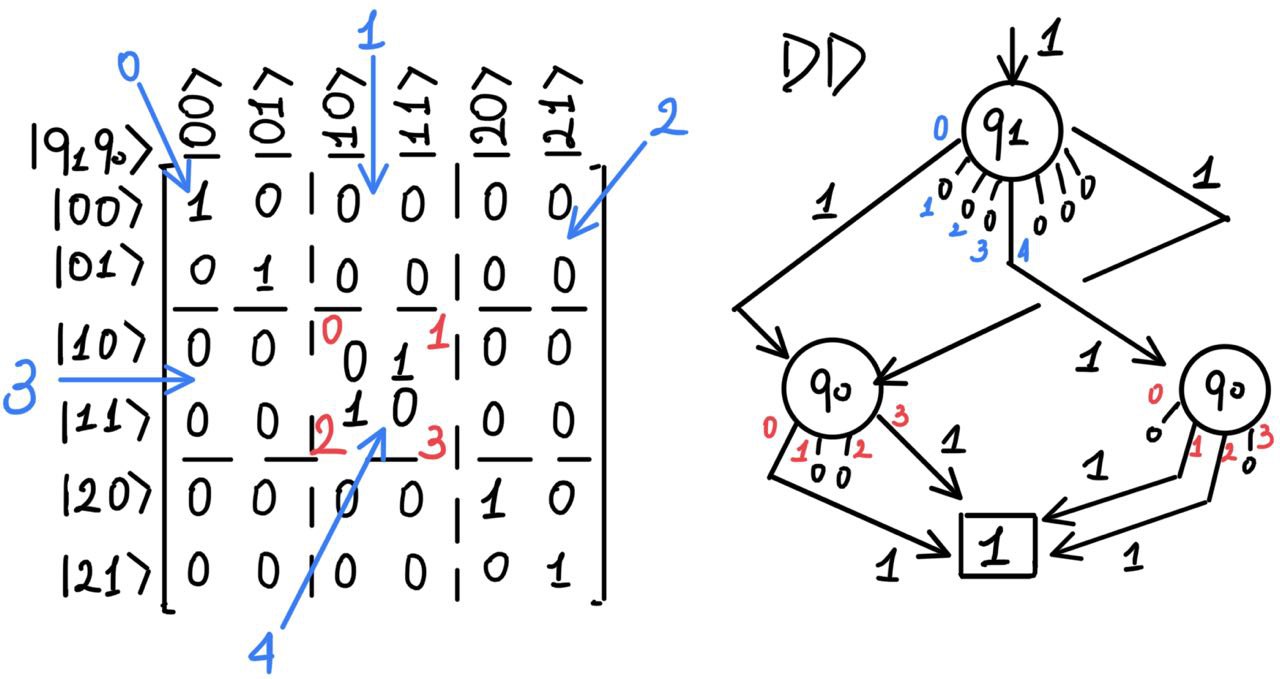}
    \caption{A unitary matrix applying the a \qop{CEX}, controlled on the level \ket{1} of a qutrit applied to \ket{0} and \ket{1} of a qubit. 
    }
    \label{fig:unitarydd}
\end{figure}

\begin{example}
    \autoref{fig:unitarydd} depicts the matrix $U$ for a \mbox{controlled-on-1-\qop{NOT}} between a qutrit and a qubit, consequently the matrix has dimension $6 \times 6$. 
    This operation, like most controlled single-qudit operations, is sparse and has high redundancy leading to a very compact decision diagram. 
    Further, the \mbox{all-zero} sub-matrices are immediately represented by \emph{zero stubs} from the root node. 
    Therefore, only three out of the nine edges of the root node representing the qutrit point to a non-zero successor. 
    The first and last edge of the root node, point to the same node that represents the pattern for the identity matrix, while the other non-zero edge in between point to a node that represents the qubit Pauli-\qop{X}.
    
    Similarly to the decision diagrams for quantum states, the individual elements of the matrix are retrieved by following a path and multiplying the complex edge weights on this path.
    Considering the element for mapping $\ket{00}$ to itself (i.e.,~the element in the upper left corner), we multiply $1 \cdot 1 \cdot 1$, corresponding to the weights of the edge to the root node and following the left-most edges from the following nodes.
\end{example}

\subsection{Rationale for Choosing DDs}

Decision diagrams have proven to be an efficient and compact data structure for exponentially-sized information in many cases in classical design automation and they have shown promising result for design automation in the quantum domain thus far.
They exhibit multiple advantages, especially for quantum circuit simulation, such as the following:
\begin{itemize}
    \item Sub-matrices that contain all-zero elements are represented with a zero-stub, i.e.,~a single edge of weight zero.
    \item Decision diagrams are canonical and, therefore, guarantee exploiting redundancies in the data structure.
    \item During operations like multiplication and addition shared nodes might be encountered more than once. This enables caching of results and, therefore, an improved run time.
    \item The Kronecker product between two decision diagrams requires only the reassignment of pointers from the terminal node of the first DD to the root node of the second (possibly followed by renormalization).
    \item Decision diagrams nicely model the locality of a single qudit.
        The number of edges at each levels matches the dimensionality of the qudit. 
\end{itemize}

The aforementioned advantages of decision diagrams make them promising candidates for quantum circuit simulation as considered in this paper.
There exist other data structures, most notably tensor networks, that are commonly employed in simulation.
However, they serve a different purpose.
For tensor networks, the strength is calculating a single observable rather than the full quantum state, where they decay to exponentially sized vectors anyway~\cite{DDvsTN}.

\section[Implementation of Mixed-Dimensional Decision Diagrams]{Implementation of\\*Mixed-Dimensional Decision Diagrams}  
\label{sec:implementation}

This section details important aspects of the implementation of the previously presented decision diagrams.
More precisely, we cover construction of decision diagrams and illustrate how the data structure is exploited for performing the basic operations for simulating quantum operations, i.e.,~multiplication and state measurements.
The full source code is available at \href{https://github.com/cda-tum/MiSiM}{github.com/cda-tum/MiSiM}.

\subsection{Representation of States and Operations}

As discussed previously, decision diagrams promise to compactly represent quantum states and quantum operations in many cases.
This, so far, rather abstract discussion has to be accompanied by an efficient implementation that can handle the representations without intermediate steps that involve the full vectors or matrices.

Instead of constructing decision diagrams representing quantum states from the corresponding vector (which require exponential memory), their direct construction is limited to basis states~\cite{DBLP:books/daglib/0046438}.
In the presented implementation we consider the basis states.
This kind of quantum state can be described linearly in the number of qudits.
For the construction of the decision diagram, one node for each qudit is required.
Building the state from a bottom-up approach will encode the state of each qudit in a single node, connected via the edges (whose edge weights encode the concrete state).
By convention, the initial quantum state in simulation is the all-zero state, which has the nodes in the decision diagram only connected by the left-most edges.

\begin{algorithm}[t]
    \caption{Make Gate DD}
    \label{alg:gatedd}
    \begin{algorithmic}
        \Require Matrix Operation $U$, lines $L$, target line $t$, controls $c \subset C$
        \Ensure root edge $e$ of Gate DD

        \State $e_{U}$ =  $U$ entries as edges
        \For{ $l$ in $L$, $l<t$ }
        \State Allocate $e_L$ = $l.size()$ zero edges
        
            \For{ $e_i$ in $e_{U}$}
                \If{$l$ is ctrl, w/ ctrl $c$ and $e_i$ is diag}
                    \State $e_{L_c} \gets e_i$
                    \For{ $e_{L_j}$ in $e_L$, $j \ne  c$}
                        \If{$e_{L_j}$ is diag} $e_{L_j} \gets I$
                        \Else $e_{L_j} \gets$ zero
                        \EndIf
                    \EndFor
                \ElsIf{$l$ is ctrl, w/ ctrl $c$ and $e_i$ not diag}
                \State $e_{L_c} \gets e_i$
                \Else
                    \Comment{Line is not a control}
                    \For{ $e_{L_j}$ in $e_L$}
                        \State $e_{L_j} \gets e_i$
                    \EndFor
                \EndIf
            \EndFor
            \State $e_i \gets e$ to norm node w/ successors $e_L$ 
        \EndFor
        
        \State $e_{U} \gets e$ to norm node w/ successors $e_i$  \Comment{Target line}

        \For{$l$ in $L$, $l>t$} \Comment{Variables above the target}
            \State Allocate $e_L$ = $l.size()$ zero edges
            \For{ $e_j$ in $e_{L}$}
                \If{$l$ is ctrl, w/ ctrl $c$ and $e_j$ is diag}
                    \If{$j=c$} $e_j \gets e_U$
                    \Else $e_j \gets I$
                    \EndIf
                \Else \Comment{Line is not a control}
                    \If{$e_j$ is diag } $e_j \gets e_U$
                    \EndIf
                \EndIf
            \EndFor
            \State $e_{U} \gets e$ to norm node w/ successors $e_L$ 
        \EndFor
        
   \end{algorithmic}
\end{algorithm}

The second key ingredient is the efficient construction of decision diagrams representing quantum operations: the procedure for the construction is shown in \autoref{alg:gatedd}.
For the proposed simulator, we considered local operations with arbitrary controls where the matrix for the local operation is the only one that is ever actually kept in memory.
In the construction, each qudit is again considered in a bottom-up fashion, i.e.,~from $q_0$ to $q_{n-1}$.
Each qudit is either (i) the target qudit, (ii) a control qudit, or (iii) not part of the considered operation.
For the target, a node with edges holding the values of the local operation is created. 
For control qudits, if the edges that represent a mapping where the operation is applied are pointing towards the operations, the remaining edges will point to the identity operation.
This is known in the process of the construction.
However, unlike qubit systems where controls are limited to $\ket{0}$ and $\ket{1}$, the controls can be on arbitrary levels, therefore constructing controlled operations becomes increasingly complex. 
For qudits that are not part of the operation, the corresponding edges represent the identity operation on these qudits.


To further optimize the construction process, special patterns in the sub-matrices are identified. 
These patterns include the identity matrix, symmetric, and transposed matrices. 
Especially for the identity matrix, sub-graphs are stored in a lookup table and referenced within the decision diagram when needed. 
This approach avoids the repeated construction of identity operations and speeds up normalization routines. 
Additionally, caching identity patterns helps to efficiently construct decision diagrams for rotations of sub-spaces within the Hilbert space, where matrices mostly consist of ones with only a few entries representing the actual operation. 

The key difference to previous implementations of decision diagrams is the added capability of dynamically creating decision diagrams with different dimensions for each qudit.
More precisely from the implementation perspective, the number of out-edges of a node can be changed to accommodate the given dimensionality.
This marks a significant departure from the fixed and static structures of previous types of decision diagrams~\cite{DBLP:conf/date/AbdollahiP06,DBLP:journals/ieicet/WangLTK08,DBLP:books/daglib/0027785,Miller2006,DBLP:journals/tcad/NiemannWMTD16,DBLP:conf/iccad/ZulehnerHW19}, making it adaptable for future use cases where temporary expansion of the Hilbert space of a qudit could be a crucial component for shorter circuits. 
The implemented data structure is closer to the behavior of nature than ever before.


\vspace{4pt}
\subsection{Applying Quantum Operations}

Given the construction of quantum states and operations detailed in the previous section, this section covers the part of the implementation that applies the operations to the states.
To this end, we consider the Kronecker product, multiplication of decision diagrams, and measurement as the essential parts of the implementation.

\subsubsection{Kronecker Product}

The first part we consider for applying quantum operations to states is the Kronecker product for decision diagrams.
For matrices, it is defined as
\begin{align*}
    A\otimes B = 
    \begin{bmatrix}
    a_{0,0}\cdot B & \cdots & a_{0,D-1}\cdot B \\
    \vdots & \ddots & \vdots \\
    a_{D-1,0} \cdot B & \cdots & a_{D-1,D-1} \cdot B
    \end{bmatrix}.
\end{align*}
This means that each element of $A$ is replaced by the element $a_{i,j} \cdot B$. 
Moreover, on matrices the Kronecker product is computationally expensive, since each of the elements of the exponentially-sized matrices has to be accessed at least once.

\begin{figure}[tpb]
    \centering
    \includegraphics[width=0.7\linewidth]{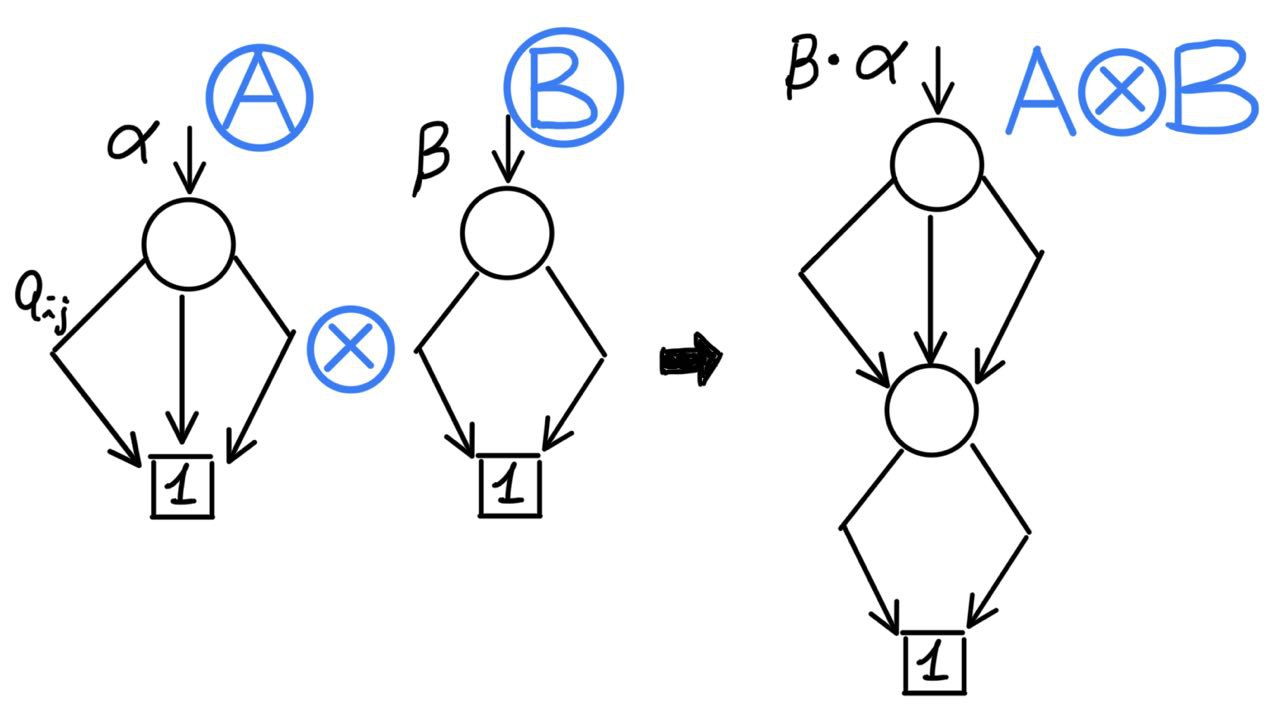}
    \caption{Kronecker product for decision diagrams}
    \label{fig:kron}
\end{figure}

However, for decision diagrams, the Kronecker product is efficient.
The terminal node of a decision diagram representing~$A$ is replaced by the root node of $B$ by means of changing the corresponding pointers.
Afterwards, for normalization, the weight of the previous edge pointing to the root node of $B$ is multiplied to the root edge to $A$.
This process is detailed in \autoref{alg:kron} and illustrated in \autoref{fig:kron}.

\begin{algorithm}[t]
	\begin{algorithmic}
	\Require root edge $x$, root edge $y$
		\Ensure root edge $e$ of $x \otimes y$ 
		\If {$x$ or $y$ is zero} \Return zero \EndIf
		\If {$x$ is scalar } \Return $y.weight \times x.weight$ \EndIf
		\If {$LookUp(x,y)$} \Return $cached(x,y)$ \EndIf
		\State Allocate $e_N$ = $x.node.size()$ zero edges 
		
		\For{$e_i$ in $e_N$}
		    \State $e_i \gets \textit{Kron}(x.node.e_i,y)$
		\EndFor
		\State $e_K \gets e$ to norm node w/ successors $e_N$ 
		\State $e_K.weight \gets e_K.weight \times x.weight \times y.weight$
		\State\Return $e_K$
	\end{algorithmic}
	\caption{Kronecker Product}\label{alg:kron}
\end{algorithm}


The implementation of the Kronecker product uses a recursive approach that leverages the decision diagram; the computational complexity of the operation is linear in the nodes of the left-hand side operand. 

\subsubsection{Multiplication and Addition}

As discussed in \autoref{sec:simulation}, multiplication and addition are the quintessential operations necessary to conduct quantum circuit simulation.
Given a quantum state $\ket{\psi}$ (with vector representation $\psi$) and a unitary operation $U$, the operation is applied by:
\begin{align*}
    \psi'_i = \sum_{k=0}^{D-1} u_{i,k}\cdot \psi_k,
\end{align*}
where $D$ is the dimension of the vector and $u_{i,k}$ ($\psi_k)$ represents an element in the matrix (vector).
We can use the recursive conceptual decomposition of the matrix (vector) represented by the respective DD, but this time to perform the multiplication. We use
\begin{align*}
    \psi' = U\cdot \psi & = 
    \begin{bmatrix}
            U_{0,0} & \cdots & U_{0,D-1} \\
            \vdots & \ddots & \vdots \\
            U_{D-1,0}  & \cdots & U_{D-1,D-1}
    \end{bmatrix} \cdot
    \begin{bmatrix}
        \psi_0 \\
        \vdots \\
        \psi_{D-1}
    \end{bmatrix}
    \\
    & = \begin{bmatrix}
        \sigma_0 \\
        \vdots \\
        \lambda_0
    \end{bmatrix} + \cdots +
    \begin{bmatrix}
        \sigma_{D-1} \\
        \vdots \\
        \lambda_{D-1}
    \end{bmatrix}
\end{align*}
The algorithm will recursively follow corresponding paths in the DDs for the matrix and the vector, and apply the operation of multiplication between the sub-matrices and sub-vectors, as detailed in \autoref{alg: mult} and \autoref{alg:add}. All the intermediate state vectors can be added recursively, i.e., 
\begin{align*}
       \psi'= 
        \begin{bmatrix}
        \sigma_0 + \cdots +\sigma_{D-1} \\
        \vdots \\
        \lambda_0 + \cdots +\lambda_{D-1}
    \end{bmatrix}
\end{align*}
The order and structure of the \mbox{sub-multiplications} and \mbox{sub-additions} follow the structure of the decision diagram.
Consequently, the complexity is bound to the decision diagram in terms of number of nodes.
However, the reduction of the nodes allows to cache intermediate results. Finally, the reduction of DDs potentially delivers an exponential improvement in terms of memory complexity, as well as in terms of time complexity by terminating early sub-computations already performed. 
The final step of multiplication is, naturally, the normalization of the decision diagram.
The procedure is recursive and included in the steps described in \autoref{alg:add}.

The multiplication and addition operations have been optimized to account for the dynamic structure of the decision diagrams. Unlike previous versions of the data structure where the number of edges was fixed and known beforehand (i.e., pointers in an array), the current version uses a vector with \mbox{variable-sized} edges (pointers), which allows for greater flexibility. Despite the added overhead of checking the size and iterating over each edge, the implementation remains relatively efficient compared to previous fixed versions of the data structure.

\begin{algorithm}[t]
    \begin{algorithmic}
    \Require root edge $x$, root edge $y$
    \Ensure root edge $e$ of $x \times y$ 
    
    \If {$x$ or $y$ is NULL or $x.weight$ or $y.weight = 0$} \Return zero \EndIf
    \If {$Terminal(x)$ and $Terminal(y)$} \Return Terminal $e$, w/ $e.weight = x.weight \times y.weight$ \EndIf
    \If {$LookUp(x,y)$} \Return $cached(x,y)$ \EndIf
    \If {$x$ is $I$} \Return $y$ \EndIf
    \If {$y$ is $I$} \Return $x$ \EndIf
    
    \State Allocate $e_N[R]$ zero edges \Comment{Edges for Results}
    
    \For{$i$ in $R$} \Comment{Rows}
    \For{$j$ in $C$} \Comment{Columns}
    \For{$e_k$ in $e_N$, $k = R * i + j$;}
        \State $e_k \gets e_k + Mult(e_x[R \cdot i + k], e_y[j + C \cdot k]$)
    \EndFor
    \EndFor
    \EndFor
    
    \State $Cache \ e_K \gets e$ to norm node w/ successors $e_N$ 
    \State $e_K.weight \gets e_K.weight \times x.weight \times y.weight$ 
    \State \Return $e_K$
    \end{algorithmic}
    \caption{Multiplication} \label{alg: mult}
\end{algorithm}

\begin{algorithm}[t]
    \begin{algorithmic}
    \Require root edge $x$, root edge $y$
    \Ensure root edge $e$ of $x + y$ 
    
    \If {$Terminal(x)$} \Return $y$ \EndIf
    \If {$Terminal(y)$} \Return $x$ \EndIf
    \If {$x.node = y.node$} \Return $e$, w/ $e.weight = x.weight+y.weight$ and $e.node=x.node$ \EndIf
    \If {$LookUp(x,y)$} \Return $cached(x,y)$ \EndIf
    \State Create $e_N$ edges with $ e_N.size() = x.nodes.edges.size$ 
    
    \For{$i = 0\ldots e_N.size()-1$} 
        \State $e_{y,next}\gets e_{yi.node};e_{y,next.weight} \gets y_{weight}{\times} e_{y_i.weight}$
        \State $e_{x,next} \gets e_{xi.node};e_{x,next.weight} \gets x_{weight}{\times} e_{x_i.weight}$
        \State $e_j = add(e_{x,next},e_{y,next}), i=j$
    \EndFor
    \State \Return $e$ to norm node w/ successors $e_N$
    \end{algorithmic}
    \caption{Addition} \label{alg:add}
\end{algorithm}

\subsubsection{Measurement}
The measurement can be interpreted as a global measurement since the outcome of the simulation is the state vector of the system.
Decision diagrams enable a measurement procedure that is linear in the number of qudits.

The process of measuring (or sampling) from the decision diagram requires one traversal of the decision diagram in a top-to-bottom manner, starting from the root node.
At each node, the squared magnitude of each edge weight are the probability that this edge should be followed.
After following the edge to the next node, this procedure is repeated until the target node is reached.
The corresponding basis state is then given by the path through the decision diagram (the index of each edge gives the corresponding level, starting with zero) and the product of the edge weights on the path gives the corresponding amplitude.

\section{Experimental Evaluation}
\label{sec:results}

\begin{table*}[tp]
    \centering
    \caption{Evaluations}
    \label{tab:evaluations}
    \renewcommand{\arraystretch}{1.2}
    \begin{tabular}{@{}lrrrrrrrS[table-format=9.0]S[table-auto-round=true,table-format=5.3]@{}}
    	\toprule
    	Benchmark      & \#Qudits & \multicolumn{4}{c}{\#Qudits with Dimension} & \#Operations & \#Nodes & {\#Distinct $\mathbb{C}$} & {Run time}   \\
    	\cmidrule{3-6} &          &  2 &   3 & 4 &                        5 &              &         &                           &              \\ \midrule
    	Mixed W-State  &        4 &  2 &   2 & 0 &                        0 &           15 &      12 & 9                         & 0.000218     \\
    	               &       30 & 15 &   3 & 0 &                       12 &           96 &      60 & 64                        & 0.016963     \\
    	               &       54 &  2 &  52 & 0 &                        0 &          159 &     108 & 112                       & 0.026502     \\
    	               &       60 &  8 &   4 & 0 &                       48 &          213 &     120 & 140                       & 0.169142     \\
    	               &       90 & 15 &  63 & 0 &                       12 &          276 &     180 & 198                       & 0.103204     \\
    	               &       90 &  2 &  80 & 0 &                        8 &          273 &     180 & 195                       & 0.100824     \\
    	               &      102 & 75 &   2 & 0 &                       25 &          315 &     200 & 207                       & 0.111326     \\
    	               &      108 &  8 & 100 & 0 &                        0 &          321 &     216 & 229                       & 0.102217     \\[4pt]
    	GHZ State      &        5 &  0 &   5 & 0 &                        0 &            8 &      14 & 5                         & 0.000089     \\
    	               &       10 &  0 &  10 & 0 &                        0 &           18 &      29 & 5                         & 0.000329     \\
    	               &       30 &  0 &  30 & 0 &                        0 &           58 &      89 & 5                         & 0.002814     \\
    	               &       60 &  0 &  60 & 0 &                        0 &          118 &     179 & 5                         & 0.010835     \\
    	               &      120 &  0 & 120 & 0 &                        0 &          238 &     359 & 5                         & 0.043214     \\
    	               &      128 &  0 & 128 & 0 &                        0 &          254 &     383 & 5                         & 0.049299     \\[4pt]

                     Random          &         2 &       0 &   0 & 2 &                        0 &        2000 &      6 &     68903 &   0.021342 \\
                                    &         2 &       0 &   2 & 0 &                        0 &        2000 &      5 &     40835 &  0.0145411 \\
                                    &         2 &           0 &   0 & 0 &                        2 &        2000 &      7 &    112918 &  0.0677329 \\
                                    &         2 &        0 &   1 & 1 &                        0 &         2000 &       5 & 52451      & 0.015843     \\
                                    &         2 &    0 &  1 & 1 &                       0 &         2000 &       5 & 52451                     & 0.016498     \\[2pt]
                                    
                                    &         3 &      2 &   1 & 0 &                        0 &        3000 &     10 &     66575 &  0.0212878 \\
                                    &         3 &      0 & 0 & 2 &1 &        3000 &     27 &    531995 &   0.171818 \\
                                    &         3 &      0 & 0 & 2 &1 &        3000 &     27 &    545173 &   0.184389 \\
                                    &         3 &      1 & 0 & 2 & 0 &        3000 &     12 &    168915 &  0.0483138 \\
                                    &         3 &      1 & 0 & 2 & 0 &        3000 &     14 &    227513 &  0.0685854 \\[2pt]
                                    &         4 &     1 & 2 & 0 &1 &        4000 &     67 &    774865 &   0.338785 \\
                                    &        4 &  1 &   1 & 1 & 1 &         4000 &      44 & 870383   & 0.380040   \\
                                    &         4 &     2 & 2 & 0 &0 &        4000 &     22 &    239247 &  0.0691839 \\
                                    &         4 &     0 & 1 & 1 &2 &        4000 &    107 &   2651118 &    4.17189 \\
                                    &         4 &     2 & 1 & 0 &1 &        4000 &     41 &    429614 &   0.152886 \\
                                    &         4 &     0 & 0 & 2 &2 &        4000 &    106 &   3141654 &     6.5643 \\
                                    &         4 &     0 & 2 & 1 &1 &        4000 &     53 &   1299120 &   0.828366 \\[2pt]
                                    &         5 &    0 & 0 & 3 &2 &        5000 &    507 &  16771562 &    231.361 \\
                                    &         5 &    1 & 1 & 2 &1 &        5000 &    214 &   4578367 &    13.6223 \\
                                    &         5 &    1 & 1 & 2 &1 &        5000 &    161 &   3287282 &    6.59248 \\
                                    &         5 &    2 & 0 & 1 &2 &        5000 &    137 &   3181180 &    6.25842 \\
                                    &         5 &    2 & 1 & 0 &2 &        5000 &    101 &   2319028 &    3.09375 \\
                                    &         5 &    2 & 0 & 1 &2 &        5000 &    164 &   3435001 &    7.24165 \\
                                    &         5 &    1 & 1 & 1 & 2&         5000 &     217 & 5732896 & 22.777900    \\[2pt]
    	                              &        6 &  2 &   1 & 1 & 2 &         6000 &     434 & 11781434 & 105.238000   \\
                                    &         6 &   2 & 2 & 1 &1 &        6000 &    486 &   7518639 &     38.225 \\
                                    &         6 &   2 & 1 & 0 &3 &        6000 &    437 &  13889217 &    159.516 \\
                                    &         6 &   4 & 2 & 0 &0 &        6000 &    137 &   1324892 &   0.958793 \\[2pt]
                                    &         7 &  3 & 1 & 2 &1 &        7000 &   1308 &  23540630 &    390.539 \\
                                    &         7 &  3 & 1 & 0 &3  &        7000 &   2447 &  42457590 &     1275.1 \\
                                    &         7 &  3 & 2 & 0 &2 &        7000 &    577 &  16793344 &      227.6 \\
                                    &        7 &  2 & 2 & 1 &2 &         7000 &    1301 & 41679061                  & 1407.100000  \\[2pt]
                                    &        8 & 5 & 2& 1 & 0                   & 8000& 572& 11114401       &        85.3092 \\
                                    &        8 &  2 &   3 & 1 & 2 &         8000 &    3902 & 135137166 &         14972.200000 \\
                                    &        8 & 2& 1& 2 & 3                    & 8000& 19886& 432117869    &      170405 \\
                    \bottomrule
    \end{tabular}
    
    \raggedright\vspace{3pt}
    The column \enquote{\#Distinct $\mathbb{C}$} shows the number of different complex numbers in the decision diagram.
\end{table*}
We implemented the proposed approach and evaluated its applicability.
The implementation is publicly available at \href{https://github.com/cda-tum/MiSiM}{github.com/cda-tum/MiSiM} as part of the Munich Quantum Toolkit (MQT,~\cite{DBLP:conf/giqc/WilleHB22}).

The evaluation was performed on a server running GNU/Linux with an Intel Xeon W-1370P (running at \SI{3.6}{\giga\hertz}) and \SI{128}{\gibi\byte} main memory.
The implementation is written in C++ and was compiled with GCC~9.3.0.
To demonstrate the capabilities, a set of three different algorithms with multiple instances was considered more precisely:
\begin{itemize}
    \item Mixed W-States~\cite{yeh2023scaling} given as states $\frac{1}{\sqrt{n}}(\ket{0\ldots01} + \ket{0\ldots010} + \ket{10\ldots0})$. Qu\emph{b}it W-states embedded in prime-dimensional qu\emph{d}its.
    \item \emph{GHZ States}~\cite{greenberger2007going} were considered for $n$ qutrits as $\frac{1}{\sqrt{3}}(\ket{0}^{\otimes n} + \ket{1}^{\otimes n} + \ket{2}^{\otimes n})$.
    \item Random Circuits built from randomly selected local operations (Hadamard and Givens rotations) and entangling operations (\qop{CEX} and controlled Clifford operations).
\end{itemize}

The results are summarized in \autoref{tab:evaluations}.
The first column lists the name of the benchmark.
The second column gives the total number of qudits where as the following four columns give the number of qudits for each level, i.e.,~the number of qubits~(2), qutrits (3), ququads (4), and ququints (5).
In the following three columns, the number of operations, number of nodes, and number of distinct complex numbers in the decision diagram of the final state after the simulation are listed.
Finally, the last column lists the run time of each instance of the benchmark.

For the algorithms Mixed W-State and GHZ state, \autoref{tab:evaluations} nicely confirms the efficiency of the proposed approach and the corresponding implementation.
The simulation for these algorithms completes in less than a second with a low number of nodes and distinct complex numbers, as one might expect.
In contrast, the random circuits contain little to no redundancy to be exploited by the decision diagrams and, therefore, can be considered to show the worst-case behavior.
Still, the implementation can simulate these circuits with, for example, 8000~operations on 8~qudits within around \SI{4}{\hour}, which, given the individual number of levels, are comparable to 14 qubits. The simulator is remarkable in its ability to deliver reliable results for randomized circuits, also when those are comparable to a larger 15-qubit system. Although the simulation process largely terminates within two days, this time frame is a testament to the efficiency and accuracy of the simulator's algorithms.
These results confirm the efficacy of the simulator as a design tool for mixed-dimensional quantum circuits.

\section{Conclusions}
\label{sec:conclusions}

Design automation is an essential component in the development of quantum computers and a key factor in unlocking their full potential.
In this paper, we proposed a classical simulator for mixed-dimensional qudit systems. To this end, we introduced a type of decision diagram capable of handling the simulation of mixed-dimensional quantum circuits and presented a corresponding implementation (publicly available at \href{https://github.com/cda-tum/MiSiM}{github.com/cda-tum/MiSiM}).
More precisely, the proposed decision diagrams encode the dimensionality of a qudit in the number of out-going edges of a node, enabling a dynamic handling of the dimension and fast adaption to the restrictions imposed, e.g.,~after calibration of a qudit system.
This compact representation is the basis for the corresponding implementation, which enables utilization of the method in real world context.
The experimental evaluation confirms the efficacy for the selected set of benchmarks, ranging from easy circuits such as the GHZ state to random circuits, which show the worst-case behavior.


\section*{Acknowledgment}
{\small This work has received funding from the European Union’s Horizon 2020 research and innovation programme under the ERC Consolidator Grant (agreement No~101001318) and the NeQST Grant (agreement No~101080086).
It is part of the Munich Quantum Valley, which is supported by the Bavarian state government with funds from the Hightech Agenda Bayern Plus and was partially supported by the BMK, BMDW, and the State of Upper Austria in the frame of the COMET program (managed by the FFG).}
\eject
\printbibliography

@STRING{tcad	= {{IEEE} Trans. on {CAD} of Integrated Circuits and Systems} }

@STRING{ieice	= {IEICE Trans. Fundamentals} }

@STRING{siam	= {SIAM Jour. of Comp.} }

@STRING{el	= {Electronic Letters} }

@STRING{iccad	= {Int'l Conf. on CAD} }

@STRING{dac	= {Design Automation Conf.} }

@STRING{aspdac	= {Asia and South Pacific Design Automation Conf.} }

@STRING{control	= {Control Conference} }

@STRING{acs	= {Applications of Computer Systems} }

@STRING{ismvl	= {Int'l Symp. on {M}ulti-{V}alued {L}ogic} }

@STRING{stoc	= {Symp. on Theory of Computing} }

@STRING{hpec  = {High Performance Extreme Computing Conf.} }

@STRING{tqc   = {ACM Trans. on Quantum Computing} }

@STRING{qce   = {Int'l Conf. on Quantum Computing and Engineering} }

@STRING{qsw   = {Int'l Conf. on Quantum Software} }

@article{brennen2005efficient,
    author = {Brennen, G. K. and Bullock, S. S. and O'Leary, D. P.},
    title = {Efficient Circuits for Exact-Universal Computationwith Qudits},
    year = {2006},
    issue_date = {July 2006},
    publisher = {Rinton Press, Incorporated},
    volume = {6},
    number = {4},
    issn = {1533-7146},
    journal = {Quantum Info. Comput.},
    pages = {436–454},
    numpages = {19}
}

@article{Cao_2019,
	doi = {10.1021/acs.chemrev.8b00803},
	url = {https://doi.org/10.1021/acs.chemrev.8b00803},
	year = 2019,
	publisher = {American Chemical Society ({ACS})},
	volume = {119},
	number = {19},
	pages = {10856--10915},
	author = {Yudong Cao and Jonathan Romero and Jonathan P. Olson and Matthias Degroote and Peter D. Johnson and M{\'{a}
}ria Kieferov{\'{a}} and Ian D. Kivlichan and Tim Menke and Borja Peropadre and Nicolas P. D. Sawaya and Sukin Sim and Libor Veis and Al{\'{a}}n Aspuru-Guzik},
	title = {Quantum Chemistry in the Age of Quantum Computing},
	journal = {Chemical Reviews}
}

@misc{deller2022quantum,
      title={Quantum approximate optimization algorithm for qudit systems with long-range interactions}, 
      author={Yannick Deller and Sebastian Schmitt and Maciej Lewenstein and Steve Lenk and Marika Federer and Fred Jendrzejewski and Philipp Hauke and Valentin Kasper},
      year={2022},
      eprint={2204.00340},
      archivePrefix={arXiv},
      xxprimaryClass={quant-ph}
}

@inproceedings{DBLP:conf/dac/HillmichMW20,
  author       = {Stefan Hillmich and
                  Igor L. Markov and
                  Robert Wille},
  title        = {Just Like the Real Thing: Fast Weak Simulation of Quantum Computation},
  booktitle    = dac,
  pages        = {1--6},
  publisher    = {{IEEE}},
  year         = {2020},
  url          = {https://doi.org/10.1109/DAC18072.2020.9218555},
  doi          = {10.1109/DAC18072.2020.9218555},
  bibsource    = {dblp computer science bibliography, https://dblp.org}
}

@article{10.1145/3491248,
author = {Jaques, Samuel and H\"{a}ner, Thomas},
title = {Leveraging State Sparsity for More Efficient Quantum Simulations},
year = {2022},
issue_date = {September 2022},
publisher = {Association for Computing Machinery},
volume = {3},
number = {3},
issn = {2643-6809},
url = {https://doi.org/10.1145/3491248},
doi = {10.1145/3491248},
journal = tqc,
articleno = {15},
numpages = {17}
}

@inproceedings{DBLP:conf/qce/MatoRHW22,
  author       = {Kevin Mato and
                  Martin Ringbauer and
                  Stefan Hillmich and
                  Robert Wille},
  title        = {Adaptive Compilation of Multi-Level Quantum Operations},
  booktitle    = qce,
  pages        = {484--491},
  publisher    = {{IEEE}},
  year         = {2022},
  url          = {https://doi.org/10.1109/QCE53715.2022.00070},
  doi          = {10.1109/QCE53715.2022.00070},
  bibsource    = {dblp computer science bibliography, https://dblp.org}
}

@inproceedings{DBLP:conf/aspdac/MatoRHW23,
  author       = {Kevin Mato and
                  Martin Ringbauer and
                  Stefan Hillmich and
                  Robert Wille},
  title        = {Compilation of Entangling Gates for High-Dimensional Quantum Systems},
  booktitle    = aspdac,
  pages        = {202--208},
  publisher    = {{ACM}},
  year         = {2023},
  url          = {https://doi.org/10.1145/3566097.3567930},
  doi          = {10.1145/3566097.3567930},
  bibsource    = {dblp computer science bibliography, https://dblp.org}
}

@INPROCEEDINGS{9926318,
  author={Stavenger, Timothy J and Crane, Eleanor and Smith, Kevin C and Kang, Christopher T and Girvin, Steven M and Wiebe, Nathan},
  booktitle=hpec, 
  title={{C2QA} - Bosonic Qiskit}, 
  year={2022},
  volume={},
  number={},
  doi={10.1109/HPEC55821.2022.9926318}
}

@book{DBLP:books/daglib/0046438,
  author       = {Michael A. Nielsen and
                  Isaac L. Chuang},
  title        = {Quantum Computation and Quantum Information (10th Anniversary edition)},
  publisher    = {Cambridge University Press},
  year         = {2016},
  url          = {https://www.cambridge.org/de/academic/subjects/physics/quantum-physics-quantum-information-and-quantum-computation/quantum-computation-and-quantum-information-10th-anniversary-edition?format=HB},
  isbn         = {978-1-10-700217-3},
  bibsource    = {dblp computer science bibliography, https://dblp.org}
}

@article{shortcuts,
author = {Lanyon, Ben and Barbieri, M. and Almeida, Marcelo and Jennewein, Thomas and Ralph, T. and Resch, K. and Pryde, G. and O'Brien, Jeremy and Gilchrist, Alexei and White, Andrew},
year = {2008},
pages = {},
title = {Quantum computing using shortcuts through higher dimensions},
volume = {5},
journal = {Nature Physics},
doi = {10.1038/nphys1150.}
}

@misc{DDvsTN,
      title={Tensor Networks or Decision Diagrams? {Guidelines} for Classical Quantum Circuit Simulation}, 
      author={Lukas Burgholzer and Alexander Ploier and Robert Wille},
      year={2023},
      eprint={2302.06616},
      archivePrefix={arXiv},
      xxprimaryClass={quant-ph}
}

@inproceedings{grover1996fast,
  author       = {Lov K. Grover},
  editor       = {Gary L. Miller},
  title        = {A Fast Quantum Mechanical Algorithm for Database Search},
  booktitle    = stoc,
  pages        = {212--219},
  publisher    = {{ACM}},
  year         = {1996},
  url          = {https://doi.org/10.1145/237814.237866},
  doi          = {10.1145/237814.237866},
  bibsource    = {dblp computer science bibliography, https://dblp.org}
}

@article{shor,
	doi = {10.1137/s0097539795293172},
  
	url = {https://doi.org/10.1137%2Fs0097539795293172},
  
	year = 1997,
  
	publisher = {Society for Industrial {\&} Applied Mathematics ({SIAM})},
  
	volume = {26},
  
	number = {5},
  
	pages = {1484--1509},
  
	author = {Peter W. Shor},
  
	title = {Polynomial-Time Algorithms for Prime Factorization and Discrete Logarithms on a Quantum Computer},
  
	journal = {{SIAM} Journal on Computing}
}

@misc{qudiet,
      title={{QuDiet}: {A} Classical Simulation Platform for Qubit-Qudit Hybrid Quantum Systems}, 
      author={Turbasu Chatterjee and Arnav Das and Subhayu Kumar Bala and Amit Saha and Anupam Chattopadhyay and Amlan Chakrabarti},
      year={2022},
      eprint={2211.07918},
      archivePrefix={arXiv},
      xxprimaryClass={quant-ph}
}

@article{qibo,
    doi = {10.1088/2058-9565/ac39f5},
    url = {https://dx.doi.org/10.1088/2058-9565/ac39f5},
    year = {2021},
    publisher = {IOP Publishing},
    volume = {7},
    number = {1},
    pages = {015018},
    author = {Stavros Efthymiou and Sergi Ramos-Calderer and Carlos Bravo-Prieto and Adrián Pérez-Salinas and Diego García-Martín and Artur Garcia-Saez and José Ignacio Latorre and Stefano Carrazza},
    title = {Qibo: a framework for quantum simulation with hardware acceleration},
    journal = {Quantum Science and Technology}
}

@article{yao,
	doi = {10.22331/q-2020-10-11-341},
	url = {https://doi.org/10.22331/q-2020-10-11-341},
	year = 2020,
	publisher = {Verein zur Forderung des Open Access Publizierens in den Quantenwissenschaften},
	volume = {4},
	pages = {341},
	author = {Xiu-Zhe Luo and Jin-Guo Liu and Pan Zhang and Lei Wang},
	title = {Yao.jl: Extensible, Efficient Framework for Quantum Algorithm Design},
	journal = {Quantum}
}

@misc{qsimCirq,
      title={Simulations of Quantum Circuits with Approximate Noise using {qsim} and {Cirq}}, 
      author={Sergei V. Isakov and Dvir Kafri and Orion Martin and Catherine Vollgraff Heidweiller and Wojciech Mruczkiewicz and Matthew P. Harrigan and Nicholas C. Rubin and Ross Thomson and Michael Broughton and Kevin Kissell and Evan Peters and Erik Gustafson and Andy C. Y. Li and Henry Lamm and Gabriel Perdue and Alan K. Ho and Doug Strain and Sergio Boixo},
      year={2021},
      eprint={2111.02396},
      archivePrefix={arXiv},
      xxprimaryClass={quant-ph}
}

@article{tenpy,
	doi = {10.21468/scipostphyslectnotes.5},
	url = {https://doi.org/10.21468/scipostphyslectnotes.5},
	year = 2018,
	publisher = {Stichting {SciPost}},
	author = {Johannes Hauschild and Frank Pollmann},
	title = {Efficient numerical simulations with Tensor Networks: Tensor Network  Python ({TeNPy})},
	journal = {{SciPost} Phys. Lecture Notes}
}

@article{quest,
	doi = {10.1038/s41598-019-47174-9},
	url = {https://doi.org/10.1038%2Fs41598-019-47174-9},
	year = 2019,
	publisher = {Springer Science and Business Media {LLC}},
	volume = {9},
	number = {1},
	author = {Tyson Jones and Anna Brown and Ian Bush and Simon C. Benjamin},
	title = {{QuEST} and High Performance Simulation of Quantum Computers},
	journal = {Sci. Rep.}
}

@Article{itensor,
	title={{The {ITensor} Software Library for Tensor Network Calculations}},
	author={Matthew Fishman and Steven R. White and E. Miles Stoudenmire},
	journal={SciPost Phys. Codebases},
	pages={4},
	year={2022},
	publisher={SciPost},
	doi={10.21468/SciPostPhysCodeb.4},
	url={https://scipost.org/10.21468/SciPostPhysCodeb.4},
}

@article{quimb,
  doi = {10.21105/joss.00819},
  url = {https://doi.org/10.21105/joss.00819},
  year = {2018},
   publisher = {The Open Journal},
  volume = {3},
  number = {29},
  pages = {819},
  author = {Johnnie Gray},
  title = {quimb: A python package for quantum information and many-body calculations},
  journal = {Journal of Open Source Software}
}

@inproceedings{Miller2006,
  doi = {10.1109/ismvl.2006.35},
  url = {https://doi.org/10.1109/ismvl.2006.35},
  year = {2006},
  publisher = {{IEEE}},
  author = {D.M. Miller and M.A. Thornton},
  title = {{QMDD}: A Decision Diagram Structure for Reversible and Quantum Circuits},
  booktitle = ismvl
}

@inproceedings{commtrade,
  author={Litteken, Andrew and Baker, Jonathan M. and Chong, Frederic T.},
  booktitle=ismvl, 
  title={Communication Trade Offs in Intermediate Qudit Circuits}, 
  year={2022},
  volume={},
  number={},
  pages={43-49},
  doi={10.1109/ISMVL52857.2022.00014}
}

@Article{molecularspin,
    author ="Tacchino, F. and Chiesa, A. and Sessoli, R. and Tavernelli, I. and Carretta, S.",
    title  ="A proposal for using molecular spin qudits as quantum simulators of light–matter interactions",
    journal  ="J. Mater. Chem. C",
    year  ="2021",
    volume  ="9",
    issue  ="32",
    pages  ="10266-10275",
    publisher  ="The Royal Society of Chemistry",
    doi  ="10.1039/D1TC00851J",
    url  ="http://dx.doi.org/10.1039/D1TC00851J"
}

@inproceedings{timeeff,
    author = {L. Seifert and J. Chadwick and A. Litteken and F. T. Chong and J. M. Baker},
    booktitle = qce,
    title = {Time-Efficient Qudit Gates through Incremental Pulse Re-seeding},
    year = {2022},
    volume = {},
    issn = {},
    pages = {304-313},
    doi = {10.1109/QCE53715.2022.00051},
    url = {https://doi.ieeecomputersociety.org/10.1109/QCE53715.2022.00051}
}

@article{phitheory,
    title = {Quantum simulation of $\phi^4$ theories in qudit systems},
    author = {Kurkcuoglu, Doga Murat and Alam, M. Sohaib and Li, Andy C.Y. and Macridin, Alexandru and Perdue, Gabriel N.},
    doi = {},
    url = {https://www.osti.gov/biblio/1834167}, 
    journal = {TBD},
    place = {United States},
    year = {2021},
}

@inproceedings{volyaaspdac,
author = {Volya, Daniel and Mishra, Prabhat},
title = {Quantum Data Compression for Efficient Generation of Control Pulses},
year = {2023},
isbn = {9781450397834},
publisher = {ACM},
url = {https://doi.org/10.1145/3566097.3567927},
doi = {10.1145/3566097.3567927},
pages = {216–221},
numpages = {6},
booktitle = aspdac
}

@article{ZXW,
  doi = {10.48550/ARXIV.2302.12135},
  url = {https://arxiv.org/abs/2302.12135},
  author = {Poór, Boldizsár and Wang, Quanlong and Shaikh, Razin A. and Yeh, Lia and Yeung, Richie and Coecke, Bob},
  title = {Completeness for arbitrary finite dimensions of {ZXW}-calculus, a unifying calculus},
  publisher = {arXiv},
  year = {2023},
  copyright = {arXiv.org perpetual, non-exclusive license}
}

@article{fermBoson,
author = {Lamata, Lucas and Mezzacapo, Antonio and Casanova, Jorge and Solano, Enrique},
year = {2014},
pages = {9},
title = {Efficient quantum simulation of fermionic and bosonic models in trapped ions},
volume = {1},
journal = {EPJ Quantum Technology},
doi = {10.1140/epjqt9}
}

@article{quditbasedQED,
  title = {Prospects for simulating a qudit-based model of $(1+1)\mathrm{D}$ scalar QED},
  author = {Gustafson, Erik J.},
  journal = {Phys. Rev. D},
  volume = {103},
  issue = {11},
  pages = {114505},
  numpages = {11},
  year = {2021},
  publisher = {American Physical Society},
  doi = {10.1103/PhysRevD.103.114505},
  url = {https://link.aps.org/doi/10.1103/PhysRevD.103.114505}
}

@article{nativequdit,
    author="Hrmo, Pavel
    and Wilhelm, Benjamin
    and Gerster, Lukas
    and van Mourik, Martin W.
    and Huber, Marcus
    and Blatt, Rainer
    and Schindler, Philipp
    and Monz, Thomas
    and Ringbauer, Martin",
    title="Native qudit entanglement in a trapped ion quantum processor",
    journal="Nature Communications",
    year="2023",
    day="19",
    volume="14",
    number="1",
    pages="2242",
    issn="2041-1723",
    doi="10.1038/s41467-023-37375-2",
    url="https://doi.org/10.1038/s41467-023-37375-2"
}

@Article{ringbauer2021universal,
	author="Ringbauer, Martin
	and Meth, Michael
	and Postler, Lukas
	and Stricker, Roman
	and Blatt, Rainer
	and Schindler, Philipp
	and Monz, Thomas",
	title="A universal qudit quantum processor with trapped ions",
	journal="Nature Physics",
	year="2022",
	day="01",
	volume="18",
	number="9",
	pages="1053--1057",
	issn="1745-2481",
	doi="10.1038/s41567-022-01658-0",
	url="https://doi.org/10.1038/s41567-022-01658-0"
}

@INPROCEEDINGS{mato2023compression,
	AUTHOR    = {K. Mato and S. Hillmich and R. Wille},
	TITLE     = {{Compression of Qubit Circuits: Mapping to Mixed-Dimensional Quantum Systems}},
	BOOKTITLE = qsw,
	YEAR      = {2023},
	ACKS      = {erc,mqv,neqst,scch},
	DOMAIN    = {quantum,quantum-map}
}

@article{DBLP:journals/tcad/BurgholzerW21,
  author       = {Lukas Burgholzer and
                  Robert Wille},
  title        = {Advanced Equivalence Checking for Quantum Circuits},
  journal      = tcad,
  volume       = {40},
  number       = {9},
  pages        = {1810--1824},
  year         = {2021},
  url          = {https://doi.org/10.1109/TCAD.2020.3032630},
  doi          = {10.1109/TCAD.2020.3032630},
  bibsource    = {dblp computer science bibliography, https://dblp.org}
}

@inproceedings{DBLP:conf/qce/HillmichHRMW21,
  author       = {Stefan Hillmich and
                  Charles Hadfield and
                  Rudy Raymond and
                  Antonio Mezzacapo and
                  Robert Wille},
  title        = {Decision Diagrams for Quantum Measurements with Shallow Circuits},
  booktitle    = qce,
  pages        = {24--34},
  publisher    = {{IEEE}},
  year         = {2021},
  url          = {https://doi.org/10.1109/QCE52317.2021.00018},
  doi          = {10.1109/QCE52317.2021.00018},
  bibsource    = {dblp computer science bibliography, https://dblp.org}
}

@Inbook{Wille2023,
    author="Wille, Robert
    and Hillmich, Stefan
    and Burgholzer, Lukas",
    title="Decision Diagrams for Quantum Computing",
    bookTitle="Design Automation of Quantum Computers",
    year="2023",
    publisher="Springer",
    pages="1--23",
    isbn="978-3-031-15699-1",
    doi="10.1007/978-3-031-15699-1_1",
    url="https://doi.org/10.1007/978-3-031-15699-1_1"
}

@article{Zhang2013Contextuality,
    author = {Zhang, Xiang and Um, Mark and Zhang, Junhua and An, Shuoming and Wang, Ye and Deng, Dong-ling and Shen, Chao and Duan, Lu-Ming and Kim, Kihwan},
    doi = {10.1103/PhysRevLett.110.070401},
    journal = {Phys. Rev. Lett.},
    pages = {070401},
    title = {State-Independent Experimental Test of Quantum Contextuality with a Single Trapped Ion},
    volume = {110},
    year = {2013}
}

@article{Lanyon2008,
    author = {Lanyon, Benjamin P. and Barbieri, Marco and Almeida, Marcelo P. and Jennewein, Thomas and Ralph, Timothy C. and Resch, Kevin J. and Pryde, Geoff J. and O'Brien, Jeremy L. and Gilchrist, Alexei and White, Andrew G.},
    doi = {10.1038/nphys1150},
    journal = {Nature Physics},
    pages = {134--140},
    title = {{Simplifying quantum logic using higher-dimensional Hilbert spaces}},
    volume = {5},
    year = {2008}
}

@article{Ringbauer2017Coherence,
    author = {Ringbauer, Martin and Bromley, Thomas R. and Cianciaruso, Marco and Lami, Ludovico and Lau, W. Y. Sarah and Adesso, Gerardo and White, Andrew G. and Fedrizzi, Alessandro and Piani, Marco},
    doi = {10.1103/PhysRevX.8.041007},
    journal = {Phys. Rev. X},
    pages = {041007},
    title = {Certification and Quantification of Multilevel Quantum Coherence},
    volume = {8},
    year = {2018}
}

@article{Hu2018a,
    author = {Hu, Xiao-Min and Guo, Yu and Liu, Bi-Heng and Huang, Yun-Feng and Li, Chuan-Feng and Guo, Guang-Can},
    doi = {10.1126/sciadv.aat9304},
    journal = {Sci. Adv.},
    number = {7},
    title = {{Beating the channel capacity limit for superdense coding with entangled ququarts}},
    volume = {4},
    year = {2018}
}

@article{Kononenko2020,
  title={Characterization of control in a superconducting qutrit using randomized benchmarking},
  author={Kononenko, M and Yurtalan, MA and Ren, S and Shi, J and Ashhab, S and Lupascu, A},
  journal={Phys. Rev. Res.},
  volume={3},
  number={4},
  pages={L042007},
  year={2021},
  publisher={APS}
}

@article{Morvan2020,
    author = {Morvan, A. and Ramasesh, V. V. and Blok, M. S. and Kreikebaum, J. M. and O'Brien, K. and Chen, L. and Mitchell, B. K. and Naik, R. K. and Santiago, D. I. and Siddiqi, I.},
    doi = {10.1103/PhysRevLett.126.210504},
    journal = {Phys. Rev. Lett.},
    pages = {210504},
    title = {Qutrit Randomized Benchmarking},
    volume = {126},
    year = {2021}
}

@article{Ahn2000,
    author = {Ahn, J. and Weinacht, T.C. and Bucksbaum, P.H.},
    doi = {10.1126/science.287.5452.463},
    journal = {Science},
    number = {5452},
    pages = {463--465},
    title = {Information Storage and Retrieval Through Quantum Phase},
    volume = {287},
    year = {2000}
}

@article{Malik2016,
    author = {Malik, Mehul and Erhard, Manuel and Huber, Marcus and Krenn, Mario and Fickler, Robert and Zeilinger, Anton},
    doi = {10.1038/nphoton.2016.12},
    journal = {Nature Photonics},
    pages = {248--252},
    title = {{Multi-photon entanglement in high dimensions}},
    volume = {10},
    year = {2016}
}

@article{Godfrin2017,
    author = {Godfrin, C. and Ferhat, A. and Ballou, R. and Klyatskaya, S. and Ruben, M. and Wernsdorfer, W. and Balestro, F.},
    doi = {10.1103/PhysRevLett.119.187702},
    journal = {Phys. Rev. Lett.},
    pages = {187702},
    title = {Operating Quantum States in Single Magnetic Molecules: {I}mplementation of {Grover}'s Quantum Algorithm},
    volume = {119},
    year = {2017}
}

@article{Anderson2015,
    author = {Anderson, B. E. and Sosa-Martinez, H. and Riofr{\'{i}}o, C. A. and Deutsch, Ivan H. and Jessen, Poul S.},
    doi = {10.1103/PhysRevLett.114.240401},
    journal = {Phys. Rev. Lett.},
    pages = {240401},
    title = {Accurate and Robust Unitary Transformations of a High-Dimensional Quantum System},
    volume = {114},
    year = {2015}
}

@article{Gedik2015,
    author = {Gedik, Z. and Silva, I. A. and {\c{C}}akmak, B. and Karpat, G. and Vidoto, E. L. G. and Soares-Pinto, D. O. and DeAzevedo, E. R. and Fanchini, F. F.},
    doi = {10.1038/srep14671},
    journal = {Sci. Rep.},
    pages = {14671},
    title = {{Computational speed-up with a single qudit}},
    volume = {5},
    year = {2015}
}

@misc{ qiskit,
       author = {H{\'e}ctor Abraham and AduOffei and Rochisha Agarwal and Gabriele Agliardi and Merav Aharoni and Vishnu Ajith and Ismail Yunus Akhalwaya and Gadi Aleksandrowicz and Thomas Alexander and Matthew Amy and Sashwat Anagolum and Anthony-Gandon and Israel F. Araujo and Eli Arbel and Abraham Asfaw and Anish Athalye and Artur Avkhadiev and Carlos Azaustre and PRATHAMESH BHOLE and Vishal Bajpe and Abhik Banerjee and Santanu Banerjee and Will Bang and Aman Bansal and Panagiotis Barkoutsos and Ashish Barnawal and George Barron and George S. Barron and Luciano Bello and Yael Ben-Haim and M. Chandler Bennett and Daniel Bevenius and Dhruv Bhatnagar and Prakhar Bhatnagar and Arjun Bhobe and Paolo Bianchini and Lev S. Bishop and Carsten Blank and Sorin Bolos and Soham Bopardikar and Samuel Bosch and Sebastian Brandhofer and Brandon and Sergey Bravyi and Bryce-Fuller and David Bucher and Artemiy Burov and Fran Cabrera and Padraic Calpin and Lauren Capelluto and Jorge Carballo and Gin{\'e}s Carrascal and Adam Carriker and Ivan Carvalho and Rishabh Chakrabarti and Adrian Chen and Chun-Fu Chen and Edward Chen and Jielun (Chris) Chen and Richard Chen and Franck Chevallier and Kartik Chinda and Rathish Cholarajan and Jerry M. Chow and Spencer Churchill and CisterMoke and Christian Claus and Christian Clauss and Caleb Clothier and Romilly Cocking and Ryan Cocuzzo and Jordan Connor and Filipe Correa and Zachary Crockett and Abigail J. Cross and Andrew W. Cross and Simon Cross and Juan Cruz-Benito and Chris Culver and Antonio D. C{\'o}rcoles-Gonzales and Navaneeth D and Sean Dague and Tareq El Dandachi and Animesh N Dangwal and Jonathan Daniel and Marcus Daniels and Matthieu Dartiailh and Abd{\'o}n Rodr{\'\i}guez Davila and Faisal Debouni and Anton Dekusar and Amol Deshmukh and Mohit Deshpande and Delton Ding and Jun Doi and Eli M. Dow and Patrick Downing and Eric Drechsler and Marc Sanz Drudis and Eugene Dumitrescu and Karel Dumon and Ivan Duran and Kareem EL-Safty and Eric Eastman and Grant Eberle and Amir Ebrahimi and Pieter Eendebak and Daniel Egger and ElePT and Iman Elsayed and Emilio and Alberto Espiricueta and Mark Everitt and Davide Facoetti and Farida and Paco Mart{\'\i}n Fern{\'a}ndez and Samuele Ferracin and Davide Ferrari and Axel Hern{\'a}ndez Ferrera and Romain Fouilland and Albert Frisch and Andreas Fuhrer and Bryce Fuller and MELVIN GEORGE and Julien Gacon and Borja Godoy Gago and Claudio Gambella and Jay M. Gambetta and Adhisha Gammanpila and Luis Garcia and Tanya Garg and Shelly Garion and James R. Garrison and Jim Garrison and Tim Gates and Gian Gentinetta and Hristo Georgiev and Leron Gil and Austin Gilliam and Aditya Giridharan and Glen and Juan Gomez-Mosquera and Gonzalo and Salvador de la Puente Gonz{\'a}lez and Jesse Gorzinski and Ian Gould and Donny Greenberg and Dmitry Grinko and Wen Guan and Dani Guijo and Guillermo-Mijares-Vilarino and John A. Gunnels and Harshit Gupta and Naman Gupta and Jakob M. G{\"u}nther and Mikael Haglund and Isabel Haide and Ikko Hamamura and Omar Costa Hamido and Frank Harkins and Kevin Hartman and Areeq Hasan and Vojtech Havlicek and Joe Hellmers and {\L}ukasz Herok and Ryan Hill and Stefan Hillmich and Colin Hong and Hiroshi Horii and Connor Howington and Shaohan Hu and Wei Hu and Chih-Han Huang and Junye Huang and Rolf Huisman and Haruki Imai and Takashi Imamichi and Kazuaki Ishizaki and Ishwor and Raban Iten and Toshinari Itoko and Alexander Ivrii and Ali Javadi and Ali Javadi-Abhari and Wahaj Javed and Qian Jianhua and Madhav Jivrajani and Kiran Johns and Scott Johnstun and Jonathan-Shoemaker and JosDenmark and JoshDumo and John Judge and Tal Kachmann and Akshay Kale and Naoki Kanazawa and Jessica Kane and Kang-Bae and Annanay Kapila and Anton Karazeev and Paul Kassebaum and Tobias Kehrer and Josh Kelso and Scott Kelso and Hugo van Kemenade and Vismai Khanderao and Spencer King and Yuri Kobayashi and Kovi11Day and Arseny Kovyrshin and Rajiv Krishnakumar and Pradeep Krishnamurthy and Vivek Krishnan and Kevin Krsulich and Prasad Kumkar and Gawel Kus and Ryan LaRose and Enrique Lacal and Rapha{\"e}l Lambert and Haggai Landa and John Lapeyre and Joe Latone and Scott Lawrence and Christina Lee and Gushu Li and Tan Jun Liang and Jake Lishman and Dennis Liu and Peng Liu and Lolcroc and Abhishek K M and Liam Madden and Yunho Maeng and Saurav Maheshkar and Kahan Majmudar and Aleksei Malyshev and Mohamed El Mandouh and Joshua Manela and Manjula and Jakub Marecek and Manoel Marques and Kunal Marwaha and Dmitri Maslov and Pawe{\l} Maszota and Dolph Mathews and Atsushi Matsuo and Farai Mazhandu and Doug McClure and Maureen McElaney and Joseph McElroy and Cameron McGarry and David McKay and Dan McPherson and Srujan Meesala and Dekel Meirom and Corey Mendell and Thomas Metcalfe and Martin Mevissen and Andrew Meyer and Antonio Mezzacapo and Rohit Midha and Declan Millar and Daniel Miller and Hannah Miller and Zlatko Minev and Abby Mitchell and Nikolaj Moll and Alejandro Montanez and Gabriel Monteiro and Michael Duane Mooring and Renier Morales and Niall Moran and David Morcuende and Seif Mostafa and Mario Motta and Romain Moyard and Prakash Murali and Daiki Murata and Jan M{\"u}ggenburg and Tristan NEMOZ and David Nadlinger and Ken Nakanishi and Giacomo Nannicini and Paul Nation and Edwin Navarro and Yehuda Naveh and Scott Wyman Neagle and Patrick Neuweiler and Aziz Ngoueya and Thien Nguyen and Johan Nicander and Nick-Singstock and Pradeep Niroula and Hassi Norlen and NuoWenLei and Lee James O'Riordan and Oluwatobi Ogunbayo and Pauline Ollitrault and Tamiya Onodera and Raul Otaolea and Steven Oud and Dan Padilha and Hanhee Paik and Soham Pal and Yuchen Pang and Ashish Panigrahi and Vincent R. Pascuzzi and Simone Perriello and Eric Peterson and Anna Phan and Kuba Pilch and Francesco Piro and Marco Pistoia and Christophe Piveteau and Julia Plewa and Pierre Pocreau and Clemens Possel and Alejandro Pozas-Kerstjens and Rafa{\l} Pracht and Milos Prokop and Viktor Prutyanov and Sumit Puri and Daniel Puzzuoli and Pythonix and Jes{\'u}s P{\'e}rez and Quant02 and Quintiii and Rafey Iqbal Rahman and Arun Raja and Roshan Rajeev and Isha Rajput and Nipun Ramagiri and Anirudh Rao and Rudy Raymond and Oliver Reardon-Smith and Rafael Mart{\'\i}n-Cuevas Redondo and Max Reuter and Julia Rice and Matt Riedemann and Rietesh and Drew Risinger and Pedro Rivero and Marcello La Rocca and Diego M. Rodr{\'\i}guez and RohithKarur and Ben Rosand and Max Rossmannek and Mingi Ryu and Tharrmashastha SAPV and Nahum Rosa Cruz Sa and Arijit Saha and Abdullah Ash- Saki and Arfat Salman and Sankalp Sanand and Martin Sandberg and Hirmay Sandesara and Ritvik Sapra and Hayk Sargsyan and Aniruddha Sarkar and Ninad Sathaye and Niko Savola and Bruno Schmitt and Chris Schnabel and Zachary Schoenfeld and Travis L. Scholten and Eddie Schoute and Mark Schulterbrandt and Joachim Schwarm and Paul Schweigert and James Seaward and Sergi and Ismael Faro Sertage and Kanav Setia and Freya Shah and Nathan Shammah and Will Shanks and Rohan Sharma and Polly Shaw and Yunong Shi and Jonathan Shoemaker and Adenilton Silva and Andrea Simonetto and Deeksha Singh and Divyanshu Singh and Parmeet Singh and Phattharaporn Singkanipa and Yukio Siraichi and Siri and Jesus Sistos and Jes{\'u}s Sistos and Iskandar Sitdikov and Seyon Sivarajah and Slavikmew and Magnus Berg Sletfjerding and John A. Smolin and Mathias Soeken and Igor Olegovich Sokolov and Igor Sokolov and Vicente P. Soloviev and SooluThomas and Starfish and Dominik Steenken and Matt Stypulkoski and Adrien Suau and Shaojun Sun and Kevin J. Sung and Makoto Suwama and Oskar S{\l}owik and Rohit Taeja and Hitomi Takahashi and Tanvesh Takawale and Ivano Tavernelli and Charles Taylor and Pete Taylour and Soolu Thomas and Kevin Tian and Mathieu Tillet and Maddy Tod and Miroslav Tomasik and Caroline Tornow and Enrique de la Torre and Juan Luis S{\'a}nchez Toural and Kenso Trabing and Matthew Treinish and Dimitar Trenev and TrishaPe and Felix Truger and Georgios Tsilimigkounakis and Davindra Tulsi and Do{\u{g}}ukan Tuna and Wes Turner and Yotam Vaknin and Carmen Recio Valcarce and Francois Varchon and Adish Vartak and Almudena Carrera Vazquez and Prajjwal Vijaywargiya and Victor Villar and Bhargav Vishnu and Desiree Vogt-Lee and Christophe Vuillot and WQ and James Weaver and Johannes Weidenfeller and Rafal Wieczorek and Jonathan A. Wildstrom and Jessica Wilson and Erick Winston and WinterSoldier and Jack J. Woehr and Stefan Woerner and Ryan Woo and Christopher J. Wood and Ryan Wood and Steve Wood and James Wootton and Matt Wright and Lucy Xing and Jintao YU and Yaiza and Bo Yang and Unchun Yang and Jimmy Yao and Daniyar Yeralin and Ryota Yonekura and David Yonge-Mallo and Ryuhei Yoshida and Richard Young and Jessie Yu and Lebin Yu and Yuma-Nakamura and Christopher Zachow and Laura Zdanski and Helena Zhang and Iulia Zidaru and Bastian Zimmermann and Christa Zoufal and aeddins-ibm and alexzhang13 and b63 and bartek-bartlomiej and bcamorrison and brandhsn and nick bronn and chetmurthy and choerst-ibm and comet and dalin27 and deeplokhande and dekel.meirom and derwind and dime10 and dlasecki and ehchen and ewinston and fanizzamarco and fs1132429 and gadial and galeinston and georgezhou20 and georgios-ts and gruu and hhorii and hhyap and hykavitha and itoko and jeppevinkel and jessica-angel7 and jezerjojo14 and jliu45 and johannesgreiner and jscott2 and kUmezawa and klinvill and krutik2966 and ma5x and michelle4654 and msuwama and nico-lgrs and nrhawkins and ntgiwsvp and ordmoj and sagar pahwa and pritamsinha2304 and rithikaadiga and ryancocuzzo and saktar-unr and saswati-qiskit and sebastian-mair and septembrr and sethmerkel and sg495 and shaashwat and smturro2 and sternparky and strickroman and tigerjack and tsura-crisaldo and upsideon and vadebayo49 and welien and willhbang and wmurphy-collabstar and yang.luh and yuri@FreeBSD and Mantas {\v{C}}epulkovskis},
       title = {Qiskit: An Open-source Framework for Quantum Computing},
       year = {2021},
       doi = {10.5281/zenodo.2573505}
}

@article{RydbergSim,
	doi = {10.1103/physrevlett.129.160501},
  
	url = {https://doi.org/10.1103%2Fphysrevlett.129.160501},
  
	year = 2022,
  
	publisher = {American Physical Society ({APS})},
  
	volume = {129},
  
	number = {16},
  
	author = {Daniel Gonz{\'{a}
}lez-Cuadra and Torsten V. Zache and Jose Carrasco and Barbara Kraus and Peter Zoller},
  
	title = {Hardware Efficient Quantum Simulation of Non-Abelian Gauge Theories with Qudits on {Rydberg} Platforms},
  
	journal = {Physical Review Letters}
}

@misc{yeh2023scaling,
      title={Scaling {W} state circuits in the qudit {Clifford} hierarchy}, 
      author={Lia Yeh},
      year={2023},
      eprint={2304.12504},
      archivePrefix={arXiv},
      xxprimaryClass={quant-ph}
}

@Inbook{greenberger2007going,
	author="Greenberger, Daniel M.
	and Horne, Michael A.
	and Zeilinger, Anton",
	editor="Kafatos, Menas",
	title="Going Beyond Bell's Theorem",
	bookTitle="Bell's Theorem, Quantum Theory and Conceptions of the Universe",
	year="1989",
	publisher="Springer",
	pages="69--72",
	isbn="978-94-017-0849-4",
	doi="10.1007/978-94-017-0849-4_10",
	url="https://doi.org/10.1007/978-94-017-0849-4_10"
}

@article{wang2020qudits,
  title={Qudits and high-dimensional quantum computing},
  author={Wang, Yuchen and Hu, Zixuan and Sanders, Barry C and Kais, Sabre},
  journal={Frontiers in Physics},
  volume={8},
  pages={589504},
  year={2020},
  publisher={Frontiers Media SA}
}

@inproceedings{DBLP:conf/iccad/ZulehnerHW19,
  author       = {Alwin Zulehner and
                  Stefan Hillmich and
                  Robert Wille},
  title        = {How to Efficiently Handle Complex Values? {I}mplementing Decision Diagrams
                  for Quantum Computing},
  booktitle    = iccad,
  pages        = {1--7},
  publisher    = {{ACM}},
  year         = {2019},
  url          = {https://doi.org/10.1109/ICCAD45719.2019.8942057},
  doi          = {10.1109/ICCAD45719.2019.8942057},
  bibsource    = {dblp computer science bibliography, https://dblp.org}
}

@article{DBLP:journals/tcad/NiemannWMTD16,
  author    = {Philipp Niemann and
               Robert Wille and
               D. Michael Miller and
               Mitchell A. Thornton and
               Rolf Drechsler},
  title     = {{QMDDs}: Efficient Quantum Function Representation and Manipulation},
  journal   = {{IEEE} Trans. Comput. Aided Des. Integr. Circuits Syst.},
  volume    = {35},
  number    = {1},
  pages     = {86--99},
  year      = {2016}
}

@book{DBLP:books/daglib/0027785,
  author    = {George F. Viamontes and
               Igor L. Markov and
               John P. Hayes},
  title     = {Quantum Circuit Simulation},
  publisher = {Springer},
  year      = {2009}
}

@article{DBLP:journals/ieicet/WangLTK08,
  author    = {Shiou{-}An Wang and
               Chin{-}Yung Lu and
               I{-}Ming Tsai and
               Sy{-}Yen Kuo},
  title     = {An {XQDD}-Based Verification Method for Quantum Circuits},
  journal   = {{IEICE} Trans. Fundam. Electron. Commun. Comput. Sci.},
  volume    = {91-A},
  number    = {2},
  pages     = {584--594},
  year      = {2008},
  url       = {https://doi.org/10.1093/ietfec/e91-a.2.584},
  doi       = {10.1093/ietfec/e91-a.2.584},
  bibsource = {dblp computer science bibliography, https://dblp.org}
}

@inproceedings{DBLP:conf/date/AbdollahiP06,
  author    = {Afshin Abdollahi and
               Massoud Pedram},
  title     = {Analysis and synthesis of quantum circuits by using quantum decision
               diagrams},
  booktitle = {Design, Automation and Test in Europe},
  pages     = {317--322},
  year      = {2006},
  url       = {https://doi.org/10.1109/DATE.2006.244176},
  doi       = {10.1109/DATE.2006.244176},
  bibsource = {dblp computer science bibliography, https://dblp.org}
}

@inproceedings{DBLP:conf/dac/BurgholzerW20,
  author       = {Lukas Burgholzer and
                  Robert Wille},
  title        = {The Power of Simulation for Equivalence Checking in Quantum Computing},
  booktitle    = dac,
  pages        = {1--6},
  publisher    = {{IEEE}},
  year         = {2020},
  url          = {https://doi.org/10.1109/DAC18072.2020.9218563},
  doi          = {10.1109/DAC18072.2020.9218563},
  bibsource    = {dblp computer science bibliography, https://dblp.org}
}

@article{DBLP:journals/tcad/ZulehnerW19,
  author       = {Alwin Zulehner and
                  Robert Wille},
  title        = {Advanced Simulation of Quantum Computations},
  journal      = tcad,
  volume       = {38},
  number       = {5},
  pages        = {848--859},
  year         = {2019},
  url          = {https://doi.org/10.1109/TCAD.2018.2834427},
  doi          = {10.1109/TCAD.2018.2834427},
  bibsource    = {dblp computer science bibliography, https://dblp.org}
}

@inproceedings{DBLP:conf/giqc/WilleHB22,
	author    = {Robert Wille and Stefan Hillmich and Lukas Burgholzer},
	title     = {{MQT}: {The Munich Quantum Toolkit}},
	booktitle = {Gesellschaft für Informatik Quantum Computing Workshop},
	year      = {2022}
}
\end{document}